%% file: CAP-MIMO-doublecluomn.tex
\documentclass[10pt,journal,final,twocolumn,]{IEEEtran}
\IEEEoverridecommandlockouts
\usepackage{amsmath,amssymb,amsfonts}
\usepackage{cite}
\usepackage{graphicx}
\usepackage{epstopdf}

\usepackage{booktabs}
\usepackage{graphicx}
\usepackage{textcomp}
\usepackage{xcolor}
\usepackage{times}
\usepackage{subfigure}         
\usepackage{amssymb,amsmath}
\usepackage{acronym}  % make an acronym
\usepackage{balance}

\usepackage{bm}         
\usepackage{algorithm} 
\usepackage{algorithmic}

\usepackage{lipsum}
\usepackage{stfloats}

\usepackage{url}

\usepackage{mathrsfs}

\usepackage{amsmath}
\allowdisplaybreaks[4]

\newtheorem{proposition}{\bf Proposition}

\newtheorem{remark}{\bf Remark}
\newtheorem{corollary}{\bf Corollary}
\newtheorem{lemma}{\bf Lemma}

\acrodef{lis}[LIS]{large intelligent surface}%\\
\acrodef{bcd}[BCD]{block coordinate descent}%\\

\acrodef{ofdm}[OFDM]{orthogonal frequency division multiplexing}%
\acrodef{miso-ofdm}[MISO-OFDM]{multi-input single-output orthogonal frequency division multiplexing}%

\acrodef{ris}[RIS]{reconfigurable intelligent surface}%

\acrodef{irs}[IRS]{intelligent reflecting surface}%

\acrodef{qos}[QoS]{quality of service}%
\acrodef{idft}[IDFT]{inverse discrete Fourier transform}%
\acrodef{dft}[DFT]{discrete Fourier transform}%
\acrodef{cp}[CP]{cyclic prefix}%
\acrodef{csi}[CSI]{channel state information}%
\acrodef{awgn}[AWGN]{additive white Gaussian noise}%

\acrodef{qcqp}[QCQP]{quadratically constrained quadratic programming}%
\acrodef{qp}[QP]{quadratic programming}%
\acrodef{bs}[BS]{base station}%
\acrodef{qos}[QoS]{quality of service}%
\acrodef{ue}[UE]{user equipment}%
\acrodef{snr}[SNR]{signal-to-noise ratio}%
\acrodef{mmwave}[mmWave]{millimeter-wave}%
\acrodef{snr}[SNR]{signal-to-noise ratio}%
\acrodef{wdm}[WDM]{wavenumber-division multiplexing}%

\acrodef{rf}[RF]{radio frequency}%

\acrodef{cap-mimo}[CAP-MIMO]{continuous-aperture MIMO}%

\acrodef{pdm}[PDM]{pattern-division multiplexing}%

\acrodef{sinr}[SINR]{signal-to-interference-plus-noise ratio}%

\acrodef{ser}[SER]{symbol error rate}%
\acrodef{mimo}[MIMO]{multiple-input multiple-output}%

\acrodef{ace}[ACE]{adaptive cross-entropy}%
\acrodef{wsr}[WSR]{weighted sum-rate}%
\acrodef{udn}[UDN]{ultra-dense network}%
\acrodef{em}[EM]{electromagnetic}%

\def\BibTeX{{\rm B\kern-.05em{\sc i\kern-.025em b}\kern-.08em
		T\kern-.1667em\lower.7ex\hbox{E}\kern-.125emX}}

\newcommand{\paperTitleMarkboth}{Pattern-Division Multiplexing for Multi-User Continuous-Aperture MIMO}

\begin{document}
	\title{Pattern-Division Multiplexing for Multi-User Continuous-Aperture MIMO 
	}
	
	\author{{{Zijian Zhang},~\IEEEmembership{Student Member,~IEEE}, and {Linglong Dai},~\IEEEmembership{Fellow,~IEEE} \vspace*{-1em}
			%{Thomas L. Marzetta,~\IEEEmembership{Fellow,~IEEE}}
		}
		\thanks{Manuscript received 23 August, 2022; revised 25 March, 2023; accepted 3 May, 2023. Date of publication XXX XX, 2023; date of current version XXX XX, 2022. This work was supported in part by the National Key Research and Development Program of China (Grant No. 2020YFB1807201), in part by the National Natural Science Foundation of China (Grant No. 62031019), and in part by the European Commission through the H2020-MSCA-ITN META WIRELESS Research Project under Grant 956256. This paper was presented in part at the IEEE ICC'22, Gangnam-gu, Seoul, South Korea, May 16–20, 2022 \cite{Zijian'22}. {\it (Corresponding author: Linglong Dai.)}
		}
		\thanks{Zijian Zhang and Linglong Dai are with the Department of Electronic Engineering, Tsinghua University, Beijing 100084, China, and also with the Beijing National Research Center for Information Science and Technology (BNRist), Beijing 100084, China (e-mail: zhangzj20@mails.tsinghua.edu.cn, daill@tsinghua.edu.cn).		}
		\thanks{Color versions of one or more figures in this article are available at https://doi.org/10.1109/JSAC.2023.3288244.}
		\thanks{Digital Object Identifier 10.1109/JSAC.2023.3288244}
	}

	\markboth{IEEE Journal on Selected Areas in Communications}{Zhang {\it et al.}: \paperTitleMarkboth}

	\maketitle
	\begin{abstract}
		In recent years, thanks to the advances in meta-materials, the concept of continuous-aperture MIMO (CAP-MIMO) is reinvestigated to achieve improved communication performance with limited antenna apertures. Unlike the classical MIMO composed of discrete antennas, CAP-MIMO has a quasi-continuous antenna surface, which is expected to generate any current distribution (i.e., pattern) and induce controllable spatial electromagnetic (EM) waves. In this way, the information is directly modulated on the EM waves, which makes it promising to approach the ultimate capacity of finite apertures. The pattern design is the key factor to determine the communication performance of CAP-MIMO, but it has not been well studied in the literature. In this paper, we develop pattern-division multiplexing (PDM) to design the patterns for CAP-MIMO. Specifically, we first study and model a typical multi-user CAP-MIMO system, which allows us to formulate the sum-rate maximization problem. Then, we develop a general PDM technique to transform the design of the continuous pattern functions to the design of their projection lengths on finite orthogonal bases, which can overcome the challenge of functional programming. Utilizing PDM, we further propose a block coordinate descent (BCD) based pattern design scheme to solve the formulated sum-rate maximization problem. Simulation results show that, the sum-rate achieved by the proposed scheme is higher than that achieved by benchmark schemes, which demonstrates the effectiveness of the developed PDM for CAP-MIMO.
	\end{abstract}
	%, which prevents most existing beamforming schemes with full CSI knowledge from practical adopting
	%\vspace{2mm}
	\section{Introduction}
	From 3G to 5G, the system performance of wireless communications has been greatly improved by the wide use of \ac{mimo} \cite{Jeffrey'14,Federico'14,Andrews'16}. Equipped with multiple discrete antennas with half-wavelength spacing, \ac{mimo} is capable of enhancing the wireless transmissions by exploiting spatial multiplexing and diversity \cite{Geoffrey'14,Yiwei'20,Chenghao'21}. Inspired by the potential benefits of \ac{mimo} with an increasing number of antennas, exploring the ultimate transmission performance of a limited \ac{mimo} aperture has attracted extensive attention in the communication community. As an ultimate \ac{mimo} structure with extremely dense antennas, the concept of \ac{cap-mimo}, which is also called as holographic \ac{mimo} \cite{Demir'21,Ghermezcheshmeh'21,Marzetta'20,Decarli'21,Sanguinetti'21}, \ac{lis} \cite{Davide'20,Pizzo'21,Hu'18,Yuan'20}, or \ac{ris} \cite{HuangHu'20,ZiyiWang'22,Ziwei'21}, is reinvestigated for wireless communications in recent years. 
	
	%In recent years, some advances in electromagnetic propagation theory have shown that, deploying sub-wavelength antennas more densely in a limited antenna aperture is able to achieve higher channel capacity \cite{Wonseok'13,Pizzo'21,HuangHu'20}, which is promising to break the performance limit of conventional \ac{mimo} with limited antenna apertures \cite{HuangHu'20}. 
	\begin{figure}[!t]
		\centering
		\includegraphics[width=3in]{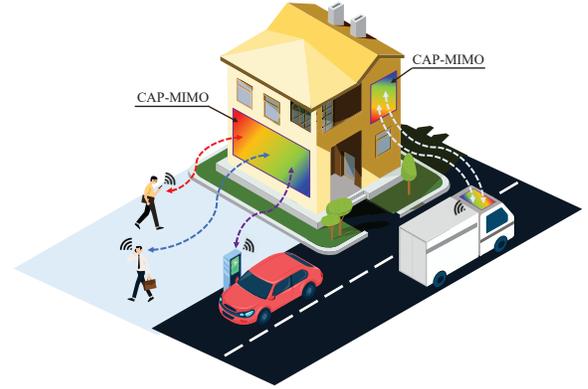}
		%\vspace{-1em}
		\caption{An example of \ac{cap-mimo} based communication scenario.}
		\label{img:good}
		%\vspace{-1em}
	\end{figure}
	Unlike the classical \ac{mimo} composed of multiple discrete antennas with half-wavelength spacing \cite{Jeffrey'14,Federico'14,Andrews'16}, by deploying a large number of sub-wavelength elements in a compact space, \ac{cap-mimo} takes the form of a quasi-continuous \ac{em} surface \cite{HuangHu'20}. Thanks to the recent advances of highly-flexible reconfigurable antennas, some early attempts at reconfigurable quasi-continuous apertures have been made \cite{Minatti'11,Hunt'14,Ovejero'17,Yurduseven'18}. For example, the authors in \cite{hwang2020binary} proposed a design method for synthesizing the binary meta-hologram pattern implemented in a leaky waveguide that can radiate signals towards a prescribed direction. Besides, by densely deploying a large number of small metamaterial elements based on electrical resonators, the research in \cite{badawe2016true} reported a continuous-aperture metasurface to achieve high array gain with limited size. In addition, the authors in \cite{hu2022arbitrary} realized a electric-driven metasurface, and the \ac{em} wave with arbitrary polarization can be generated and radiated. 
	
	As shown in Fig. \ref{img:good}, an ideal \ac{cap-mimo} is expected to have full control freedom of generating any current distribution on its spatially-continuous surface \cite{Marzetta'20}, so that its radiated \ac{em} waves can be artificially configured in a desired manner. In this way, the information for receivers can be directly modulated on the spatial \ac{em} waves and radiated to the physical space. Relying on this mechanism, the physical properties of spatial \ac{em} waves can be sufficiently exploited \cite{Wonseok'13}, leading to extreme spatial resolution \cite[Fig. 5]{Davide'20}, high spectrum efficiency \cite[Fig. 7]{Pizzo'21} \cite[Fig. 4]{Hu'18} and high energy efficiency \cite[Fig. 4]{HuangHu'20}. Thus, \ac{cap-mimo} becomes a promising technology to satisfy many challenging requirements of future wireless networks, such as the wide in-building coverage, high-speed uplink transmission, and high-accuracy localization \cite{HuangHu'20}.

	\subsection{Prior Works}
	Realizing an ideal \ac{cap-mimo} has a long history in the research field of micro-wave and photonics, dating back to Harold Wheeler’s work in 1965 \cite{Wheeler'65} and David Staiman’s work in 1968 \cite{Staiman'68}, respectively. By deriving the eigenfunctions of continuous-space \ac{em} channels, the capacity bound between two continuous volumes was derived in \cite{miller2000communicating,jensen2008capacity}. The recent works on \ac{cap-mimo} include antenna design \cite{badawe2016true}, physical model \cite{wei2022multi}, performance analysis \cite{Decarli'21}, channel estimation \cite{Demir'21}, and so on. For example, the authors in \cite{Wheeler'65} proposed to realize \ac{cap-mimo} by exploiting the current sheet made of tightly coupled dipole array and the monolayer metallic made of magnetic particles. Then, to characterize the propagation process of the \ac{em} waves, the authors in \cite{wei2022multi} modeled the wireless channels between \ac{cap-mimo} transceivers as Gaussian random fields. Subsequently, the author in \cite{Davide'20} derived the analytical expressions of the spatial degrees of freedom (DoFs) of \ac{cap-mimo}, and the authors in \cite{Decarli'21} further analyzed its near-field DoFs. Moreover, the authors in \cite{Demir'21} proposed to exploit the geometrical property of arrays to realize the overhead-reduced channel estimation for \ac{cap-mimo}. By utilizing special beam structures, a \ac{cap-mimo} channel estimation scheme was proposed in \cite{Ghermezcheshmeh'21}, whose training overhead and complexity do not scale with the number of elements. 
	
	The pattern, i.e., the current distribution on the continuous aperture of \ac{cap-mimo}, is the key factor determining the \ac{cap-mimo} performance \cite{Davide'20}. To support the coherent transmission of multiple data streams, it is necessary for \ac{cap-mimo} transmitter to adopt a series of distinguishable patterns to carry different symbols \cite{Decarli'21,Sanguinetti'21}. Specifically, the authors in \cite{Decarli'21} considered a near-field line-of-sight scenario, where one linear-aperture \ac{cap-mimo} transmitter serves one linear-aperture \ac{em}-wave receiver. By adopting a series of square-wave functions to generate the patterns, the \ac{em} waves carrying different symbols can be radiated towards different spatial angles. Furthermore, the authors in \cite{Sanguinetti'21} considered a similar near-field line-of-sight scenario with a couple of linear-aperture \ac{cap-mimo} transceivers. Particularly, a \ac{wdm} scheme was proposed to directly generate the patterns by Fourier basis functions. In this way, the transmitted symbols belonging to different streams are modulated on different spatial wavenumbers of radiated \ac{em} waves and transmitted, which is similar to the frequency-division multiplexing in conventional communications.
	
	From the above discussions, we can find that, most existing works have directly adopted the patterns generated by given special functions to realize coherent transmission for \ac{cap-mimo} \cite{Decarli'21,Sanguinetti'21}. Although these schemes can improve the performance of \ac{cap-mimo} to some extent, they only work efficiently in some special communication scenarios, such as single-user, linear-aperture, and/or near-field transmissions. To support \ac{cap-mimo} in general scenarios with complex propagation environment and multiple distributed receivers, it is essential to design the patterns flexibly according to the continuous channel functions, which resembles the space-division multiplexing in conventional \ac{mimo} systems. Unfortunately, to the best of our knowledge, such a flexible and general pattern design scheme has not been well studied in the literature. One possible reason may be the mathematical challenge introduced by the design of continuous pattern functions of \ac{cap-mimo}, which is usually non-convex functional programming \cite{Regan'16} and thus difficult to be solved by the classical discrete signal processing techniques for conventional \ac{mimo} systems.

	\subsection{Our Contributions}
	To fill in this gap, in this paper\footnote{Simulation codes are provided to reproduce the results presented in this article: http://oa.ee.tsinghua.edu.cn/dailinglong/publications/publications.html.}, we develop a general \ac{pdm} technique to flexibly design patterns for multi-user \ac{cap-mimo}. Our contributions are summarized as follows.
	\begin{itemize}
		\item Based on the \ac{em} propagation principle, we study and model a typical multi-user \ac{cap-mimo} based communication system, where one \ac{cap-mimo} transmitter with planar aperture serves multiple users coherently. This allows us to formulate the sum-rate maximization problem to optimize the \ac{cap-mimo} patterns, and it also provides a possible framework for other technical problems in multi-user \ac{cap-mimo} systems, such as channel estimation and energy efficiency optimization.
		
		\item We develop a general \ac{pdm} technique to flexibly design the patterns of \ac{cap-mimo} according to the knowledge of continuous channel functions. The key idea is to use series expansion to project the continuous pattern functions of \ac{cap-mimo} onto an orthogonal basis space, thus the design of continuous pattern functions is transformed to the design of their projection lengths on finite orthogonal bases. In this way, the challenging problem of optimizing continuous patterns of \ac{cap-mimo} can be addressed.
		
		\item Utilizing the developed \ac{pdm} technique, we further propose a \ac{bcd} based pattern design scheme to solve the formulated sum-rate maximization problem for multi-user \ac{cap-mimo}. Simulation results show that, the multi-user patterns designed by the proposed scheme are almost mutually orthogonal, and the sum-rate achieved by the proposed scheme is higher than that achieved by benchmark schemes. 
	\end{itemize}
	
	\subsection{Organization and Notation}
	\textit{Organization:} The rest of this paper is organized as follows. Section \ref{sec:sys} introduces the system model of multi-user \ac{cap-mimo} and formulates the problem of sum-rate maximization. The general \ac{pdm} technique to address the continuous patterns of \ac{cap-mimo} is developed in Section \ref{sec:Alg}, and the specific pattern design scheme to solve the formulated problem is proposed in Section \ref{sec:Method}. Simulation results are presented in Section \ref{sec:NSR} to validate the effectiveness of the proposed scheme. Finally, conclusions are drawn and future works are discussed in Section \ref{sec:con}.

	\textit{Notation:} $\mathbb{C}$, $\mathbb{R}$, $\mathbb{R}_{+}$, and $\mathbb{Z}$ denote the set of complex, real, positive real, and integer numbers, respectively; ${[\cdot]^{-1}}$, ${[\cdot]^{*}}$, ${[\cdot]^{\rm T}}$, and ${[\cdot]^{\rm H}}$ denote the inverse, conjugate, transpose, and conjugate-transpose operations, respectively;  $\|\cdot\|$ denotes the Euclidean norm of its argument; $\|\cdot\|_{\rm F}$ denotes the Frobenius norm of its argument; ${\rm det}\!\left|\, \cdot\, \right|$ denotes the determinant of its argument; $\mathbb{E}_{\bf z}\{\cdot\}$ is the expectation operator with respect to the random vector $\bf z$; $\mathfrak{R}\{\cdot\}$ denotes the real part of its argument; $\ln(\cdot)$ denotes natural logarithm; ${\rm rect}\left(\cdot\right)$ is the generalized rectangular function whose value takes one/zero when the condition in its argument is true/false; ${\rm mod}(a,b)$ denotes the remainder of $a/b$; $\nabla_{\mathbf{z}}$ denotes the first-order partial derivative operator with respect to $\bf z$; surfaces are indicated with calligraphic letters ${\cal S}$ and $\left| {{\cal S}} \right|$ denotes the Lebesgue measure of ${\cal S}$; $\mathbf{I}_{L}$ denotes an $L\times L$ identity matrix.	
	
	\section{System Model and Problem Formulation for CAP-MIMO}\label{sec:sys}
	In this section, we study and model a typical multi-user \ac{cap-mimo} based communication system, where one \ac{cap-mimo} transmitter with planer aperture simultaneously serves $K$ users in the downlink. Specifically, we first introduce the system model of a \ac{cap-mimo} transmitter in Subsection \ref{sec:sys1}. Then, the \ac{em} channels between the transmitter and the users are illustrated in Subsection \ref{sec:sys2}. Next, the \ac{em} waves at the users are modeled in Subsection \ref{sec:sys3}. Finally, the problem of the sum-rate maximization for multi-user \ac{cap-mimo} system is formulated in Subsection \ref{sec:sys4}.
	
	\subsection{\ac{cap-mimo} Transmitter}\label{sec:sys1}
	\begin{figure*}[!t]
		\centering
		\includegraphics[width=7.1in]{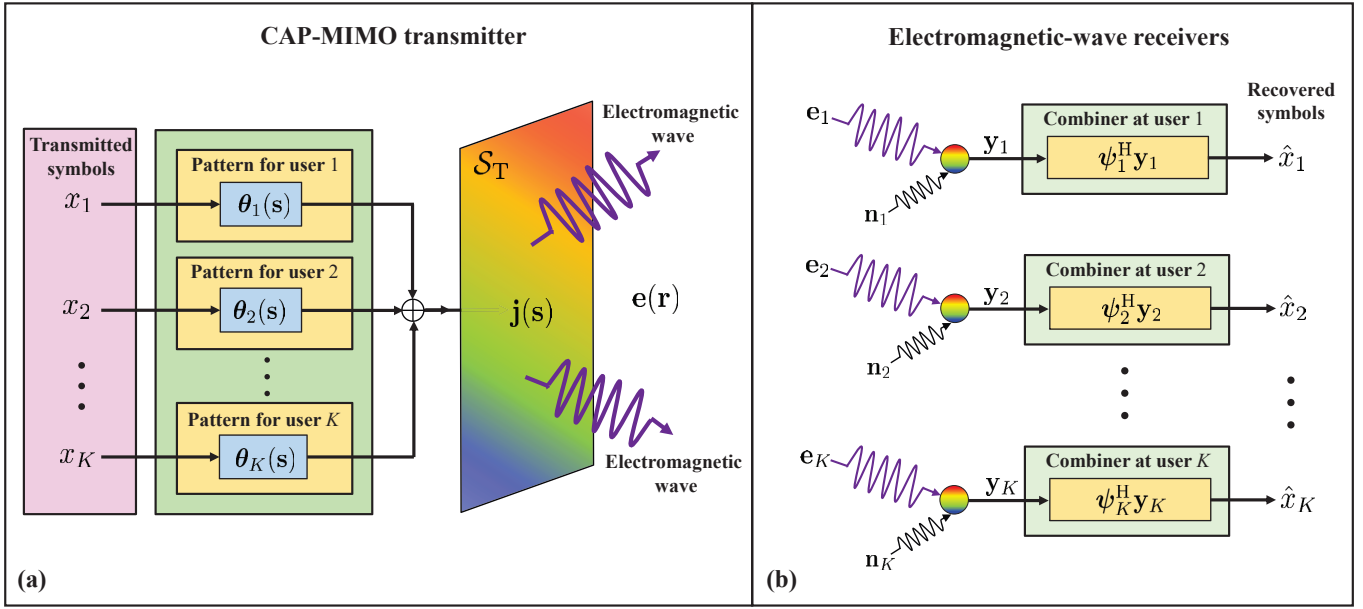}
		%\vspace{-1.5em}
		\caption{The concept of \ac{cap-mimo} based wireless communications. (a) illustrates a \ac{cap-mimo} transmitter, which has a quasi-continuous antenna aperture ${\cal S}_{\rm T}$ for radiating information-carrying \ac{em} waves ${\bf e}({\bf r})$. (b) illustrates $K$ tri-polarization single-antenna receivers (i.e., $K$ users) each being able to sense the three polarizations of incident wave ${\bf e}_k$ and then combine the three components via ${\bm \psi}_k$ to decode symbol $x_k$.}
		\label{img:CAP}
		%\vspace{-1em}
	\end{figure*}
	As shown in Fig. \ref{img:CAP} (a), we consider a \ac{cap-mimo} transmitter with aperture ${\cal S}_{\rm T}$ of area $A_{\rm T}=\left|{\cal S}_{\rm T}\right|$ working in a 3-D homogeneous medium. In the ideal case, \ac{cap-mimo} has an almost continuous antenna aperture, which is able to generate any current distribution on its continuous surface for wireless communications \cite{Marzetta'20,Decarli'21,Sanguinetti'21,Davide'20}. Let ${\bf j}({\bf s},t)\in{\mathbb R}^{3}$ denote the monochromatic current density at a generic location ${\bf s}:=\left(s_x,s_y,s_z\right)\in{\mathbb R}^{3}$ and time $t$. The ideally controllable current distribution at the \ac{cap-mimo} transmitter can be written as 
	\begin{equation}\label{eqn:js}
		\mathbf{j}(\mathbf{s}, t)=\Re\left\{\mathbf{j}(\mathbf{s}) e^{-\mathrm{j} 2\pi f t}\right\},~~{\mathbf{s}\in{{\cal S}_{\rm T}}},
	\end{equation}
	where $f$ is the current frequency. For simplicity but without loss of generality, we assume that the communication system works in narrowband, which is exactly the well-known and widely-used {\it time-harmonic assumption} in \ac{em} analysis \cite{Fred'08}. This allows us to ignore the time-related component $e^{-\mathrm{j} 2\pi f t}$ and focus on the time-independent current density ${\bf j}({\bf s})\in{\mathbb C}^3$.
	\par
	Consider that the \ac{cap-mimo} transmitter simultaneously serves $K$ tri-polarization receivers (i.e., $K$ users) in the downlink\footnote{
	In contrast to \cite{Hu'18,Yuan'20} where the uplink capacity of multi-user CAP-MIMO was analyzed, this paper focuses on the pattern design for a downlink multi-user CAP-MIMO system with tri-polarization transceivers. Due to the power constraint of CAP-MIMO, inter-user interference, and tri-polarization transmissions, the pattern design problem is challenging to solve, which motivates us to propose a general scheme based on optimizations.}. Let ${{\bf{x}}} \triangleq {\left[ {{x_{1}}, \cdots ,{x_{K}}} \right]^{\rm T}} \in {{\mathbb C}^K}$ denote the symbols transmitted to $K$ users, respectively. We assume that these symbols have the  normalized power, i.e., $\mathbb{E}_{\bf x}\left\{\mathbf{x}\mathbf{x}^{\rm H}\right\}=\mathbf{I}_{K}$. Then, similar to the conventional \ac{mimo} beamforming \cite{Nusrat'18}, the symbols $\bf x$ to be transmitted are modulated on $K$ different \ac{cap-mimo} patterns $\left\{ {{{\bm{\theta }}_k}\left( {\bf{s}} \right)} \right\}_{k = 1}^K$, which aims to make these symbols orthogonal at different users as much as possible, and thus high channel capacity can be achieved. For simplicity, we assume that \ac{cap-mimo} employs linear superposition to combine multiple information-carrying patterns $\left\{ {{{\bm{\theta }}_k}\left( {\bf{s}} \right)} \right\}_{k = 1}^K$ for coherent transmission, thus the combined current distribution ${\bf{j}}({\bf{s}})$ on the \ac{cap-mimo} aperture can be modeled as 
	\begin{equation}\label{eqn:js_theta}
		{\bf{j}}({\bf{s}}) = \sum\limits_{k = 1}^K {{{\bm{\theta }}_k}\left( {\bf{s}} \right){x_k}},~~{\mathbf{s}\in{{\cal S}_{\rm T}}}, 
	\end{equation}
	where pattern ${\bm{\theta }}_k\left( {\bf{s}} \right)\in{\mathbb C}^3$ is the component of current density that carries symbol $x_k$. 
	
	\subsection{Electromagnetic Channels}\label{sec:sys2}
	To model the radiated information-carrying \ac{em} waves in space, we define ${\bf e}({\bf r})\in\mathbb{C}^3$ as the electric field at point ${\bf r}:=\left(r_x,r_y,r_z\right)\in\mathbb{R}^3$, which is induced by the current distribution ${\bf j}({\bf s})$ on the \ac{cap-mimo} aperture. According to Maxwell's equations, the current distribution ${\bf j}({\bf r'})$ and the electric field ${\bf e}({\bf r'})$ satisfy the following inhomogeneous Helmholtz wave equation \cite{Fred'08}:
	\begin{equation}\label{eqn:Helmholtz}
		\nabla \times \nabla \times \mathbf{e}\left(\mathbf{r}^{\prime}\right)-\kappa_0^{2} \mathbf{e}\left(\mathbf{r}^{\prime}\right)=\mathrm{j} \kappa_0 Z_{0} \mathbf{j}\left(\mathbf{r}^{\prime}\right),
	\end{equation}
	where ${\bf r'}\in{\mathbb R}^3$ is any arbitrary point in space; $\kappa_0$ is the spatial wavenumber; and $Z_{0}$ is the intrinsic impedance of spatial medium, which is $376.73$ $\Omega$ in free space.
	
	Then, to explicitly express the relationship between the current distribution ${\bf j}({\bf s})$ at the transmitter and the electric field ${\bf e}({\bf r})$ at the receiver, Green's method \cite{Fred'08} is utilized to solve (\ref{eqn:Helmholtz}). By introducing channel function $\mathbf{G}(\mathbf{r}, \mathbf{s})\in{\mathbb C}^{3\times 3}$, the electric field ${\bf e}({\bf r})$ at point ${\bf r}$ can be induced from (\ref{eqn:Helmholtz}) as
	\begin{equation}\label{eqn:er}
		\mathbf{e}(\mathbf{r})=\int_{{\cal S}_{\rm T}} \mathbf{G}(\mathbf{r}, \mathbf{s}) \mathbf{j}(\mathbf{s}) \mathrm{d} \mathbf{s},
	\end{equation}
	where channel function $\mathbf{G}(\mathbf{r}, \mathbf{s})$ plays a role similar to the channel matrix in classical MIMO systems. From the perspective of mathematics, $\mathbf{G}(\mathbf{r}, \mathbf{s})$ is the system impulse response, i.e., Green function. In particular, $\mathbf{G}(\mathbf{r}, \mathbf{s})$ is determined by the specific wireless environment. For example, in ideal unbounded and homogeneous mediums, $\mathbf{G}(\mathbf{r}, \mathbf{s})$ is \cite{Fred'08}
	\begin{equation}\label{eqn:free-space}
		\mathbf{G}(\mathbf{r}, \mathbf{s})=\frac{\mathrm{j} \kappa_0 Z_{0}}{4 \pi} \frac{e^{\mathrm{j} \kappa_0\|\mathbf{r}-\mathbf{s}\|}}{\|\mathbf{r}-\mathbf{s}\|}\left(\mathbf{I}_3 - \frac{(\mathbf{r}-\mathbf{s})(\mathbf{r}-\mathbf{s})^{\rm T}}{\|\mathbf{r}-\mathbf{s}\|^2}
%		\frac{\nabla_{\mathbf{r}} \nabla_{\mathbf{r}}^{\mathrm{H}}}{\kappa_0^{2}}	
		\right)
	\end{equation}
	in the far field, while $\mathbf{G}(\mathbf{r}, \mathbf{s})$ is usually modeled as a stochastic process in scattering environments \cite{Pizzo'21}.

\subsection{Electromagnetic-Wave Receivers}\label{sec:sys3}
	As shown in Fig. \ref{img:CAP} (b), we assume that all $K$ \ac{em}-wave receivers (i.e., $K$ users) are located in the far-field region, and each user is equipped with an ideal tri-polarization antenna with area $A_{\rm R}=\frac{\lambda^2}{4\pi}$, which satisfies $A_{\rm R} \ll  A_{\rm T}$ \cite{Hu'18}. In this case, similar to a discrete antenna, each user can be reasonably approximated by a point in 3-D space. 
	
	Let ${\bf r}_k\in{\mathbb R}^3$ denote the 3-D location of the $k$-th user. In the ideal case, user $k$ with a tri-polarization antenna can sense the three polarizations of the \ac{em} waves reaching point ${\bf r}_k$ (i.e., holographic capability \cite{Decarli'21}). Then, the three polarization components are combined to decode symbol $x_k$ with a polarization combiner ${\bm \psi}_k\in{\mathbb C}^3$.\footnote{There are two potential ways to achieve an adjustable combiner ${\bm \psi}_k$: 1) The receiver antenna can actively adjust its polarization direction to combine the polarization components in the analog domain, i.e., polarization-adjustable reconfigurable antennas \cite{li2022geometry}. 2) The receiver can digitalize the polarization components and then combine them in the digital domain.} Thus, according to (\ref{eqn:js_theta}) and (\ref{eqn:er}), the \ac{em} wave received by user $k$ can be expressed as
	\begin{equation}\label{eqn:signal_model}
		\begin{aligned}
			{{\bf{y}}_k} = {{\bf{e}}_k} + {{\bf{n}}_k}
			=& \underbrace {{x_k}\int_{{{\cal S}_{\rm{T}}}} {{{\bf{G}}_k}({\bf{s}})} {\bm \theta _k}\left( {\bf{s}} \right){\rm{d}}{\bf{s}}}_{\text{Desired signal to user $k$}} + \\ & \underbrace {\sum\limits_{j = 1,j \ne k}^K {{x_j}\int_{{{\cal S}_{\rm{T}}}} {{{\bf{G}}_k}({\bf{s}}){\bm \theta _j}\left( {\bf{s}} \right)} {\rm{d}}{\bf{s}}} }_{\text{Interferences from other users}} + \underbrace {{{\bf{n}}_k}}_{\text{Noise}},
		\end{aligned}
	\end{equation}
	where ${{\bf{e}}_k}:=\mathbf{e}(\mathbf{r}_k)$, ${{\bf{G}}_k}({\bf s}):=\mathbf{G}(\mathbf{r}_k, \mathbf{s})$, and ${{\bf{n}}_k}\in{\mathbb C}^3$ is the \ac{em} noise at user $k$, which is produced by all incoming \ac{em} waves that are not generated by the transmitter \cite{Fred'08}. For simplicity, we follow the isotropic propagation assumption used in \cite{Sanguinetti'21}, thus ${{\bf{n}}_k}$ for all $k\in\{1,\cdots,K\}$ can be modeled as mutually independent \acp{awgn} with zero mean and variance $\sigma^2{\bf I}_3$. Our work can be easily extended to the general colored-noise case by replacing $\sigma^2{\bf I}_3$ in the following analysis with the specified $\mathbb{E}\left\{{\bf n}_k \mathbf{n}_k^{\mathrm{H}}\right\}$.
	
	\subsection{Sum-Rate Maximization Problem Formulation}\label{sec:sys4}
	Based on the above system model, in this subsection, we formulate the sum-rate maximization problem for multi-user \ac{cap-mimo}. By calculating the sum of multi-user mutual information, the sum-rate of $K$ users, can be derived from (\ref{eqn:signal_model}) as
	\begin{equation}\label{eqn:SR}
		{R_{\rm sum}} = \sum\limits_{k=1}^K {{\log _2}\det \left| {{{\bf{I}}_3} + {{\bm{\alpha }}_k}{\bm{\alpha }}_k^{\rm{H}}{\bf{J}}_k^{-1}} \right|}, 
	\end{equation}
	where ${\bm{\alpha }}_k$ and ${\bf{J}}_k$ are respectively given by 
	\begin{align}
		\label{eqn:alpha_J}
		{{\bm{\alpha }}_k} \!=\!& \int_{{{\cal S}_{\rm{T}}}} {{{\bf{G}}_k}({\bf{s}}){{\bm{\theta }}_k}\left( {\bf{s}} \right){\rm d}{\bf{s}}},\\
		{{\bf{J}}_k} \!=& \!\!\!\sum\limits_{j=1,j \ne k}^K \!\! { {\int_{{\cal S}_{\rm{T}}}\!\!\! {{{\bf{G}}_k}({\bf{s}}){{\bm{\theta }}_j}\left( {\bf{s}} \right){\rm d}{\bf{s}}} }{{\left( {\int_{{\cal S}_{\rm{T}}}\!\!\! {{{\bf{G}}_k}({\bf{s'}}){{\bm{\theta }}_j}\left( {\bf{s'}} \right){\rm d}{\bf{s'}}} } \right)}^{\!\!\!\rm{H}}}}  \!\!\!+ {\sigma ^2}{{\bf{I}}_3}. \notag
	\end{align}
	
	In practical systems, we are interested in investigating the maximum sum-rate subject to a given power constraint. By integrating the radial component of the Poynting vector over a sphere with infinite-length radius \cite{Fred'08}, we introduce the following lemma to upper-bound the total transmit power of \ac{cap-mimo} in the sense of expectation.
	
	\begin{lemma}[Transmit power constraint of multi-user \ac{cap-mimo}]
		The total transmit power of the multi-user \ac{cap-mimo} based communication systems can be upper-bounded by	the following inequality:
		\begin{equation}\label{eqn:power_cons}
			\sum\limits_{k = 1}^K {\int_{{{\cal S}_{\rm{T}}}} {{{\left\| {{{\bm{\theta }}_k}\left( {\bf{s}} \right)} \right\|}^2}{\rm{d}}{\bf{s}}} }  \le {P_{\rm{T}}},
		\end{equation}
		where $P_{\rm T}$ can be viewed as the allowable maximum ``transmit power'' of \ac{cap-mimo}, which is implicitly associated with the physical radiation power and measured in ${\rm A}^2$ (or ${\rm mA}^2$).
	\end{lemma}
	\begin{IEEEproof}
		Please see Appendix \ref{appendix:power}.
	\end{IEEEproof}
	
	By combing (\ref{eqn:SR}) and (\ref{eqn:power_cons}), the original problem of sum-rate maximization subject to the transmit power constraint can be formulated as
	\begin{subequations}\label{eqn:original_problem}
		\begin{align}
			\!\!\!\!{\cal P }_o:~~&\mathop{\max}\limits_{{{\bm \theta}}\left({\bf s}\right)}~~{R_{\rm sum}} = \sum\limits_{k=1}^K {{\log _2}\det \left| {{{\bf{I}}_3} + {{\bm{\alpha }}_k}{\bm{\alpha }}_k^{\rm{H}}{\bf{J}}_k^{-1}} \right|}, \label{eqn:objective} \\
			&~~{\rm s.t.}~~\sum\limits_{k = 1}^K {\int_{{{\cal S}_{\rm{T}}}} {{{\left\| {{{\bm{\theta }}_k}\left( {\bf{s}} \right)} \right\|}^2}{\rm{d}}{\bf{s}}} }  \le {P_{\rm{T}}}, \label{eqn:power}
		\end{align}
	\end{subequations}
	where ${\bm{\theta }}({\bf s})$ is defined as ${\bm{\theta }}({\bf s}):=\left\{ {{{\bm{\theta }}_k}\left( {\bf{s}} \right)} \right\}_{k = 1}^K$. Our goal is to maximize the sum-rate in (\ref{eqn:objective}) by appropriately designing the continuous pattern functions ${\bm{\theta }}({\bf s})$, i.e., the current distribution on the continuous aperture ${\cal S}_{\rm T}$ of \ac{cap-mimo}\footnote{Due to the difficulties of hardware implementations \cite{Minatti'11,Hunt'14,Ovejero'17,Yurduseven'18}, generating any patterns on a continuous aperture is challenging for current technologies. However, the optimized ${\bm{\theta }}({\bf s})$ can be viewed as ideal pattern designs for CAP-MIMO. In practice, these ideally designed patterns can be appropriately adjusted, such as low-resolution quantization or discrete sampling, to satisfy the hardware constraints of practical CAP-MIMO systems.
}. 
	
	\begin{remark}
	Generally, the pattern design problems such as ${\cal P }_o$ in (\ref{eqn:original_problem}) are difficult to solve. The reason is that, the continuous pattern functions ${\bm{\theta }}({\bf s})$ within integrals exist in both optimization objective and constraint in problem ${\cal P }_o$ in (\ref{eqn:original_problem}), which is actually non-convex functional programming \cite{Regan'16}. In this case, since the partial derivatives of the optimization objective with respect to ${\bm{\theta }}({\bf s})$ are difficult to obtain, the classical signal processing techniques for discrete arrays, such as gradient descent and Lagrange dual method \cite{Shen'18'1}, are hard to be adopted. Such kind of non-convex functional programming is common in the optics and micro-wave areas, especially for the design of the radiation patterns of directional antennas, and these problems are usually addressed by using commercial \ac{em} simulation software such as high frequency structure simulator (HFSS), which results in high time and space complexity. 
	\end{remark}
	
	\section{Developed Pattern-Division Multiplexing (PDM) for CAP-MIMO}\label{sec:Alg}
	
	To address the challenging optimization of continuous pattern functions as shown in problem ${\cal P }_o$ in (\ref{eqn:original_problem}), in this section, we develop \ac{pdm} technique to design CAP-MIMO patterns. The key idea is to use series expansion to expand the continuous pattern functions in an orthogonal basis space. In this way, the continuous pattern functions are projected onto finite orthogonal bases, thus the design of these functions is transformed to the design of their projection lengths in the orthogonal basis space, which makes optimizing continuous functions feasible. Specifically, in Subsection \ref{sec:method1}, we introduce the developed \ac{pdm} technique to deal with \ac{cap-mimo} patterns. Then, in Subsection \ref{sec:performance_analysis}, we provide the performance analysis of the developed \ac{pdm}.
	
	\subsection{A General Pattern-Division Multiplexing (PDM) Technique}\label{sec:method1}
	Different from the existing methods which adopt the patterns directly generated by special functions for coherent transmissions \cite{Decarli'21,Sanguinetti'21}, in this section, we develop \ac{pdm} technique to flexibly design the patterns for \ac{cap-mimo} according to the knowledge of continuous channel functions $\left\{{{{\bf{G}}_k}}({\bf{s}}) \right\}_{k = 1}^K$. Particularly, the developed \ac{pdm} aims to strengthen the desired information-carrying signals and make the \ac{em} waves carrying different symbols as much orthogonal as possible at the users. In this way, higher channel capacity is expected, which is similar to the space-division multiplexing in classical \ac{mimo} systems. 
	
	To efficiently optimize the continuous pattern functions $\left\{ {{{\bm{\theta }}_k}({\bf{s}})} \right\}_{k = 1}^K$ in problem ${\cal P }_o$ in (\ref{eqn:original_problem}), inspired by the pattern design methods in antenna theory, an intuitive idea is to use series expansion to project these continuous functions onto an orthogonal space \cite{Fred'08}. In this paper, we use Fourier bases to expand the spatial continuous pattern functions. Since the Fourier transform of spatial domain is exactly the wavenumber domain, this choice of bases allows us to better understand and analyze the pattern design from the perspective of wavenumber space. In this way, we obtain the following lemma.
	\begin{lemma}[Fourier series expansion of pattern functions]
		For an arbitrary continuous pattern function ${\bm{\theta }}_k({\bf s})\in{\mathbb C}^{3}$ defined in volume ${\bf s}:=\left(s_x,s_y,s_z\right)\in{\cal S}_{\rm T}$, if $	{\bm \theta}_k({\bf s})$ is absolutely integrable, pattern function $	{\bm \theta}_k({\bf s})
		$ to be designed can be equivalently rewritten as
		\begin{equation}
			{\bm \theta}_k({\bf s})=\sum\limits_{\bf{n}}^\infty  {{{\bf{w}}_{k,{\bf{n}}}}{\Psi _{\bf{n}}}\left( {\bf{s}} \right)},~~{\bf s} \in {\cal S}_{\rm T},
		\end{equation}
		where we define ${\bf{n}} := {\left({n_x},{n_y},{n_z}\right)}\in\mathbb{Z}^3$ and $\sum\nolimits_{\bf{n}}^\infty {} :=  \sum\nolimits_{{n_x} =  - \infty}^\infty {\sum\nolimits_{{n_y} =  - \infty}^\infty {\sum\nolimits_{{n_z} =  - \infty}^\infty {} } }$ to simplify notations. Particularly, the projection length ${{\bf{w}}_{k,{\bf{n}}}}\in{\mathbb C}^{3}$ in the wavenumber domain and the Fourier base function ${{\Psi _{\bf n}}\left( {\bf s} \right)}\in{\mathbb C}$ can be written as
		\begin{subequations}\label{eqn:16}
			\begin{align}
				{{\bf{w}}_{k,{\bf{n}}}} =& \frac{1}{\sqrt{A_{\rm T}}}\int_{{\bf{s}} \in {\cal S}_{\rm T}} {{\bm \theta}_k({\bf s}){\Psi^*_{\bf{n}}}\left( {\bf{s}} \right){\rm d}{\bf{s}}},~~{\bf{n}}\in\mathbb{Z}^3,\\
				{\Psi _{\bf{n}}}\left( {\bf{s}} \right) =& \frac{1}{\sqrt{A_{\rm T}}}{e^{ 2\pi{\rm j} \left( {\frac{{{n_x}}}{{{L_x}}}\left(s_x-\frac{L_x}{2}\right) + \frac{{{n_y}}}{{{L_y}}}{\left(s_y-\frac{L_y}{2}\right)} + \frac{{{n_z}}}{{{L_z}}}{\left(s_z-\frac{L_z}{2}\right)}} \right)}}, \notag \\&~~{\bf s}\in{\cal S}_{\rm T}, 
			\end{align}
		\end{subequations}
		where $L_x$, $L_y$, and $L_z$ denote the maximum projection lengths of volume ${\cal S}_{\rm T}$ on the $x$-, $y$-, and $z$-axis of 3-D coordinate system, respectively. The Fourier bases ${\Psi _{\bf{n}}}\left( {\bf{s}} \right)$ satisfy
		\begin{align}
			\int_{{\bf{s}} \in {{\cal S}_{\rm T}}} {{\Psi _{\bf{n}}}\left( {\bf{s}} \right)\Psi _{\bf{n}'}^*\left( {\bf{s}} \right)} {\rm{d}}{\bf{s}}  = \left\{ \begin{array}{l}
				1,~~{\bf n}' = {\bf n},\\
				0,~~{\bf n}' \ne {\bf n},
			\end{array} \right.
		\end{align}
		which guarantees the orthogonality between any two basis functions.
	\end{lemma}

Note that, different from the Fourier basis functions in \cite{Pizzo'21} which represent the spatial plane-wave channels, the basis functions in (\ref{eqn:16}) have no physical significance, which only provide functional degrees of freedom for the further optimization of pattern functions ${\bm \theta}({\bf s})$. Since their indexes ${\bf{n}} := {\left({n_x},{n_y},{n_z}\right)}\in\mathbb{Z}^3$ only take integer values, the equality $\left({2\pi n_x}/{L_x}\right)^2 + \left({2\pi n_y}/{L_y}\right)^2 + \left({2\pi n_z}/{L_z}\right)^2 = \kappa_0^2$ does not hold in most cases. Then, using the series expansion in {\it Lemma 2}, the continuous pattern functions $\left\{ {{{\bm{\theta }}_k}({\bf{s}})} \right\}_{k = 1}^K$ in problem ${\cal P }_o$ in (\ref{eqn:original_problem}) can be equivalently replaced by their discrete projection lengths $\left\{ {{{\bf{w}}_{k,{\bf{n}}}}} \right\}_{k = 1}^K$ in the wavenumber space. 

	\begin{figure*}[!t]
	\centering
	%\subfigcapskip -0.5em
	\subfigure[$f = 2.4$ GHz.]{\includegraphics[width=3.6in]{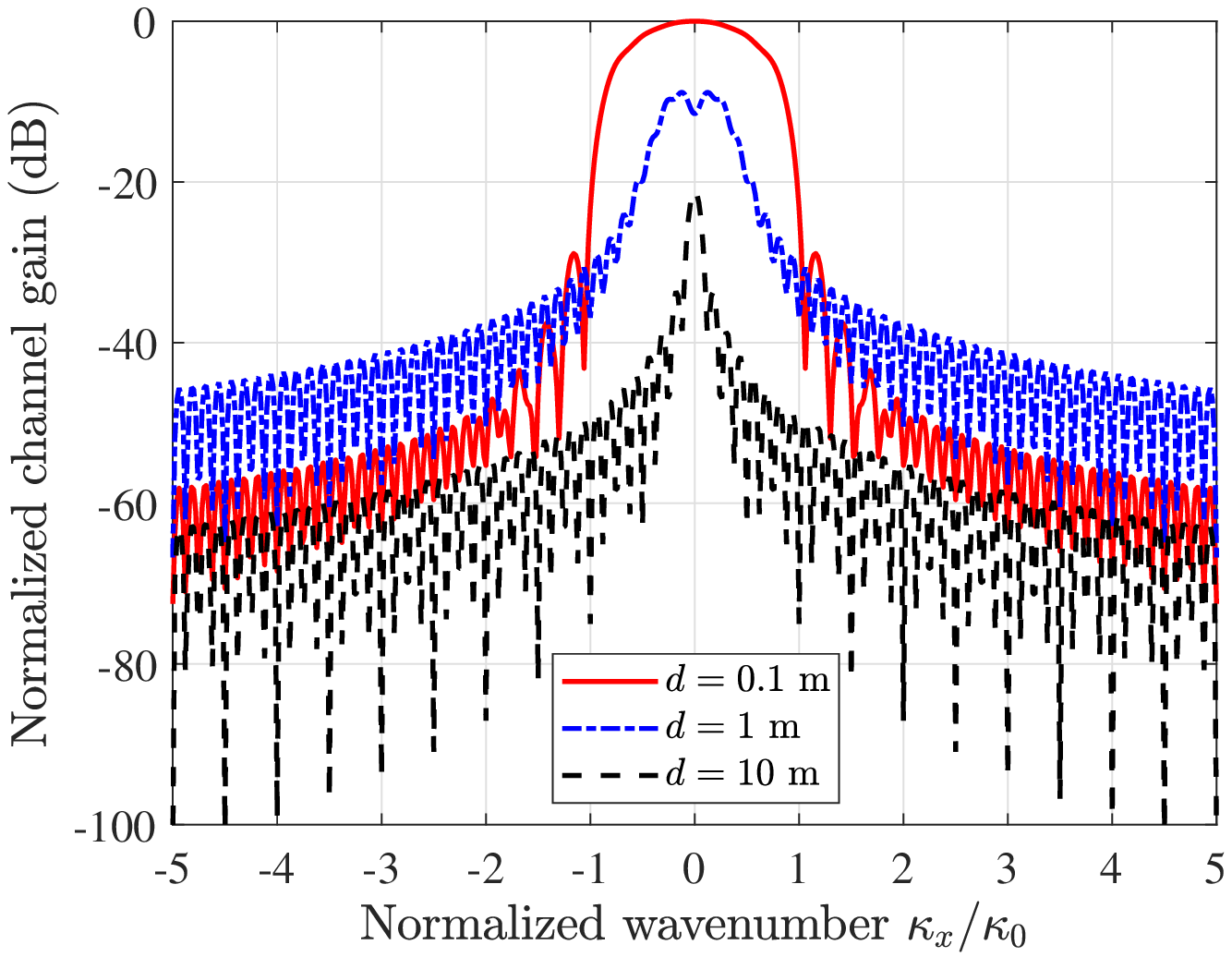}}%\hspace{-5mm}
	\subfigure[$d=0.1$ m.]{\includegraphics[width=3.6in]{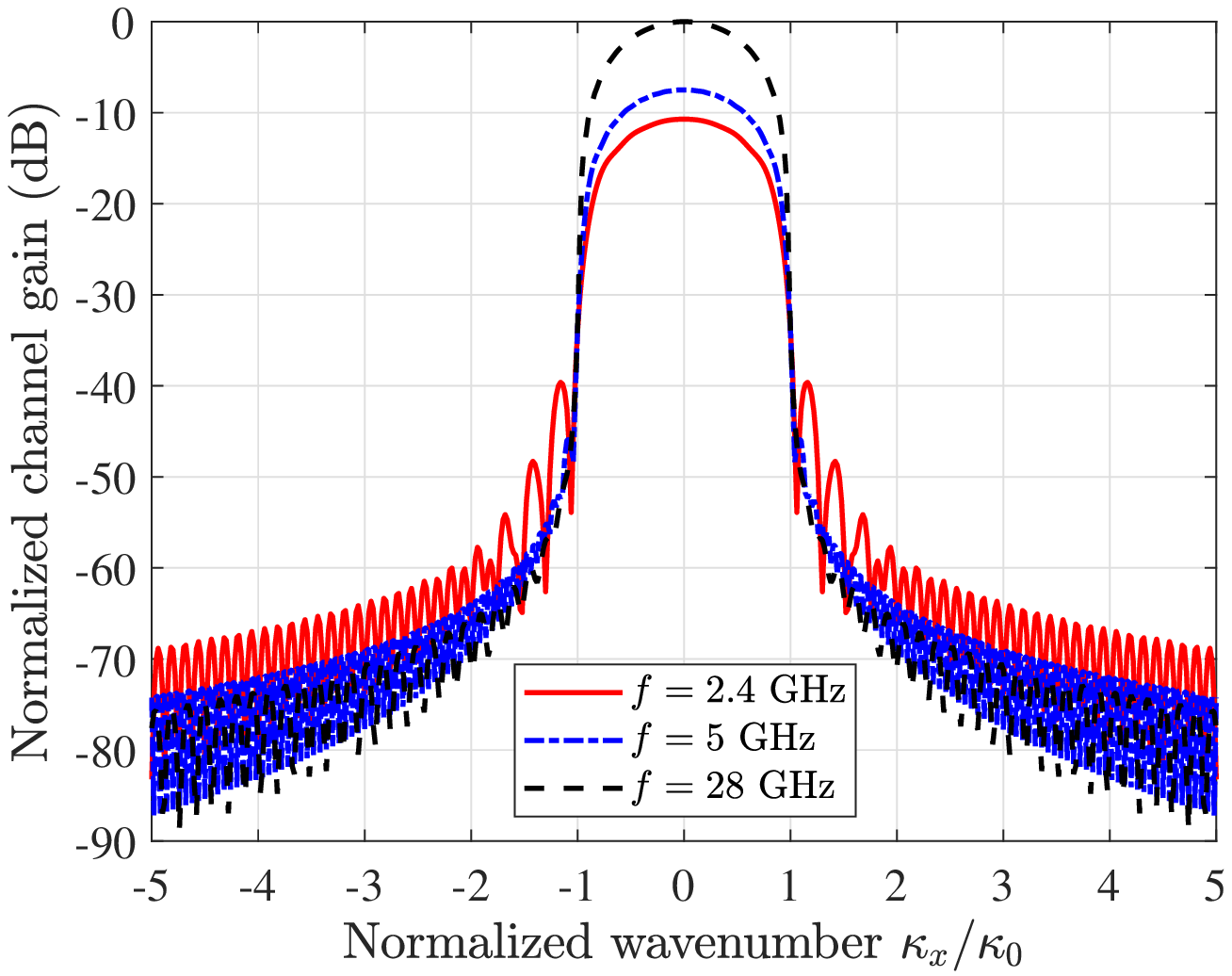}}
	%\vspace{-1em}
	\caption{The normalized channel gain $\left\| {{\mathscr F}\left\{ {{{\bf{G}}_1}({\bf{s}})}{\rm rect}\left({\bf s}\in{\cal S}_{\rm T}\right) \right\}} \right\|_{\rm{F}}^2$ against the normalized wavenumber $\kappa_x/\kappa_0$, where frequency is fixed as $f = 2.4$ GHz in (a) and distance is fixed as $d = 0.1$ m in (b).
	}
	\label{img:channel_gain}
	%\vspace{-1em}
\end{figure*}	
To this end, we introduce the following two useful corollaries to address the continuous pattern functions:
	
\begin{corollary}[Continuous-discrete transform for electric field] 
Let ${\bm{\kappa }} := \left( {{\kappa _x},{\kappa _y},{\kappa _z}} \right)$ denote the wavenumber and  ${\mathscr F}\left\{ \cdot \right\}\left( {{{\bm{\kappa }}}} \right)$ denote the Fourier transform of its argument on surface ${\cal S}_{\rm T}$ at wavenumber ${\bm{\kappa }}$. By adopting Fourier series expansion for pattern function ${\bm{\theta }}_j({\bf{s}})$, the product of channel function ${\bf{G}}_k({\bf{s}})$ and pattern function ${\bm{\theta }}_j({\bf{s}})$ in integral $\int_{{{\cal S}_{\rm T}}}{\rm d}{\bf s}$ , i.e., the electric field, can be equivalently rewritten as
	\begin{equation}\label{eqn:Green_Four}
		\begin{aligned}
			\int_{{{\cal S}_{\rm{T}}}}& {{{\bf{G}}_k}({\bf{s}}){{\bm{\theta }}_j}\left( {\bf{s}} \right){\rm d}{\bf{s}}} = \sum\limits_{\bf{n}}^\infty  {{{\bf{\Omega }}_{k,{\bf{n}}}}{{\bf{w}}_{j,{\bf{n}}}}},\\&~~~~~ \forall k,j\in\{1,\cdots,K\},
		\end{aligned}
	\end{equation}
where
	\begin{equation}
		\begin{aligned}
			{{\bf{\Omega }}_{k,{\bf{n}}}} = {\mathscr F}\left\{ {{{\bf{G}}_k}({\bf{s}})} \right\}\left( {{{\bf{\bm \kappa }}_{\bf{n}}}} \right) = \int_{{{\cal S}_{\rm{T}}}} {{{\bf{G}}_k}({\bf{s}}){\Psi _{\bf{n}}}\left( {\bf{s}} \right){\rm{d}}{\bf{s}}},
		\end{aligned}
	\end{equation}
in which ${{{\bf{\Omega }}_{k,{\bf{n}}}}}\in{\mathbb C}^{3\times 3}$ is exactly the Fourier transform of function ${\bf{G}}_k({\bf s}){\rm rect}\left({\bf s}\in{\cal S}_{\rm T}\right)$ over the \ac{cap-mimo} aperture ${\cal S}_{\rm T}$ at wavenumber ${\bm \kappa _{\bf{n}}}=(2\pi\frac{{n_x}}{L_x},2\pi\frac{{n_y}}{L_y},2\pi\frac{{n_z}}{L_z})$. In particular, according to Parseval's theorem, the following equation naturally holds:
		\begin{equation}\label{eqn:G_Omega}
			\begin{aligned}
				\int_{{{\cal S}_{\rm{T}}}} {\left\| {{{\bf{G}}_k}({\bf{s}})} \right\|_{\rm F}^2{\rm{d}}{\bf{s}}}  = \sum\limits_{\bf{n}}^\infty  {\left\| {{{\bf{\Omega }}_{k,{\bf{n}}}}} \right\|_{\rm F}^2},
			\end{aligned}
		\end{equation}
	where the left-hand side is the integral of the Frobenius norms of continuous functions while the right-hand side is the sum of the Frobenius norms of discrete matrices. Particularly, $\left\| {{{\bf{G}}_k}({\bf{s}})} \right\|_{\rm F}^2$ can be viewed as the channel gain in the spatial domain at position ${\bf r}_k$ and $\left\| {{{\bf{\Omega }}_{k,{\bf{n}}}}} \right\|_{\rm F}^2$ can be viewed as the channel gain in the wavenumber domain at wavenumber $\bm \kappa _{\bf{n}}$.
	\end{corollary}

	\begin{corollary}[Continuous-discrete transform for power constraint] 
		According to Parseval's theorem, the Euclidean norm of pattern function ${\bm{\theta }}_k\left( {\bf{s}} \right)$ in integral $\int_{{{\cal S}_{\rm T}}}{\rm d}{\bf s}$, i.e., the transmit power of $x_k$, can be equivalently rewritten as
		\begin{equation}\label{eqn:approx_theta}
			\begin{aligned}
				\int_{{{\cal S}_{\rm{T}}}} {{{\left\| {{{\bm{\theta }}_k}\left( {\bf{s}} \right)} \right\|}^2}{\rm{d}}{\bf{s}}}  = \sum\limits_{\bf{n}}^\infty  {{{\left\| {{{\bf{w}}_{k,{\bf{n}}}}} \right\|}^2}},~~ \forall k\in\{1,\cdots,K\},
			\end{aligned}
		\end{equation}
		which guarantees the conservation of system energy. 
	\end{corollary}
%	Observing continuous-discrete transforms in {\it Corollary 1} and {\it Corollary 2}, we note that the integral of the Frobenius norms of continuous functions in the left-hand side can be replaced by the sum of Frobenius norms of discrete matrices in the right-hand side.
	
	Utilizing the continuous-discrete transforms in {\it Corollary 1} and {\it Corollary 2}, we note that the continuous functions $\left\{ {{{\bf{G}}_k}}({\bf{s}}) \right\}_{k = 1}^K$ and $\left\{ {{{\bm{\theta }}_k}({\bf{s}})} \right\}_{k = 1}^K$ can be replaced by their projection lengths $\left\{ {{{\bm{\Omega }}_{k,{\bf{n}}}}} \right\}_{k = 1}^K$ and $\left\{ {{{\bf{w}}_{k,{\bf{n}}}}} \right\}_{k = 1}^K$, respectively. Thus, the functional programming \cite{Regan'16} can be reformulated as a common digital signal processing problem. However, since the number of expansion items is infinite as shown in (\ref{eqn:Green_Four}), (\ref{eqn:G_Omega}) and (\ref{eqn:approx_theta}), the optimization of projection lengths $\left\{ {{{\bf{w}}_{k,{\bf{n}}}}} \right\}_{k = 1}^K$ is still unacceptable for practical computing devices. Therefore, {\it how to address the infinite expansion items} becomes a critical issue for the pattern design of \ac{cap-mimo}

	Fortunately, thanks to the inherent physical properties of function ${\bf{G}}_k({\bf s}){\rm rect}\left({\bf s}\in{\cal S}_{\rm T}\right)$, some recent works have revealed that, the value of $\left\| {{{\bf{\Omega }}_{k,{\bf{n}}}}} \right\|_{\rm F}^2$, i.e., the channel gain in the wavenumber domain, is high in the low-wavenumber band and negligible in the high-wavenumber band in most cases. For example, the authors in \cite{Sanguinetti'21} have shown by simulations that, in a typical system with two linear-aperture \ac{cap-mimo} transceivers, the wavenumber-domain channel gain within the band of $[-\kappa_0,\kappa_0]$ is usually much higher than that within the other bands \cite[Fig. 4]{Sanguinetti'21}. By deriving the closed-form expression of ${\mathscr F}\left\{ {{{\bf{G}}}_k({\bf{s}})}{\rm rect}\left({\bf s}\in{\cal S}_{\rm T}\right) \right\}$, the authors in \cite{Zhongzhichao'21} have analytically proved that, the wavenumber-domain channel gain within the band of $[-\kappa_0,\kappa_0]$ dominates in the whole wavenumber domain  \cite[Fig. 2, Fig. 3]{Zhongzhichao'21}. Some works on channel modeling have also pointed out that, the small-scale fading of \ac{cap-mimo} channel can be modeled by the sum of finite Fourier plane-wave representations, which also implies that ignoring high-wavenumber expansion items has little impact on channel quality \cite[Eq. (39)]{Marzetta'20}.

	To observe the above fact, here we take a specific scenario for example. For simplicity, we consider a single-user \ac{cap-mimo} system working at frequency $f$, where the \ac{cap-mimo} transmitter is deployed on the x-axis with aperture ${{\cal S}_{\rm T}} = \left\{ { {\left( {{s_x},{s_y},{s_z}} \right)} {\Big |}\left| {{s_x}} \right| \le 0.5\,{\rm{m}}, {{s_y}} =0,{s_z} = 0} \right\}$ and the user is located at $\left(0,0,d\right)$. Then, we plot the normalized channel gain $\left\| {{\mathscr F}\left\{ {{{\bf{G}}_1}({\bf{s}})}{\rm rect}\left({\bf s}\in{\cal S}_{\rm T}\right) \right\}} \right\|_{\rm{F}}^2$ against the normalized wavenumber $\kappa_x/\kappa_0$ in Fig. \ref{img:channel_gain} (a) and Fig. \ref{img:channel_gain} (b) with different setups. From these two figures, one can observe that, for all our considered setups, the channel gain within the band of $\kappa_x\in[-\kappa_0,\kappa_0]$ dominates in the wavenumber domain. Particularly, when $\left| \kappa_x/\kappa_0 \right|$ is larger than 1, compared with $\left| \kappa_x/\kappa_0 \right|=0$, the wavenumber-domain channel gain suffers a large loss of about -30 dB, which is in agreement with the results in existing works \cite{Sanguinetti'21,Zhongzhichao'21,Marzetta'20}.

	As a result, this fact inspires us to approximate the Fourier expansion series $\sum\nolimits_{\bf{n}}^\infty  {{{\bf{\Omega }}_{k,{\bf{n}}}}{{\bf{w}}_{j,{\bf{n}}}}}$ with finite low-wavenumber and high-power items that satisfy $|2\pi \frac{{{n_x}}}{{{L_x}}}| \le {\kappa _0}$, $|2\pi \frac{{{n_y}}}{{{L_y}}}| \le {\kappa _0}$, and $|2\pi \frac{{{n_z}}}{{{L_z}}}| \le {\kappa _0}$ and ignore those high-wavenumber and low-power items, which actually exploits the idea of compressed sensing \cite{Donoho'06}. In this way, we obtain the following proposition.

\begin{proposition}[Finite-item approximation of expansion series] 
		Define ${\bf N}:=\left(N_x,N_y,N_z\right)$ with $N_x$, $N_y$, and $N_z$ being the numbers of the reserved expansion items on the x-, y-, and z-axis, respectively. Let $\sum\nolimits_{\bf{n}}^{\bf{N}} :  = \sum\nolimits_{{n_x} =  - {N_x}}^{{N_x}} {\sum\nolimits_{{n_y} =  - {N_y}}^{{N_y}} {\sum\nolimits_{{n_z} =  - {N_z}}^{{N_z}} {} } } $ for notation simplification. Then, the electric field in (\ref{eqn:Green_Four}) and the transmit power in (\ref{eqn:approx_theta}) can be approximated\footnote{This approximation is valid when the evanescent waves are negligible, i.e., the users are not too close to the CAP-MIMO aperture. In this case, all wavenumber components of propagating waves take the real values within $[-\kappa_0, +\kappa_0]$, thus the suggested $\bf N$ in (\ref{eqn:NxNyNz}) ensures a safe approximation.} by
		\begin{subequations}\label{eqn:truncation}
			\begin{align}
				\!\!\!\!\!\!\int_{{{\cal S}_{\rm{T}}}}\!\! {{{\bf{G}}_k}({\bf{s}}){{\bm{\theta }}_j}\left( {\bf{s}} \right){\rm d}{\bf{s}}} &\!\approx\! \sum\limits_{\bf{n}}^{\bf N}  {{{\bf{\Omega }}_{k,{\bf{n}}}}{{\bf{w}}_{j,{\bf{n}}}}}, \forall k,j\in\{1,\cdots,K\}, \\
				\int_{{{\cal S}_{\rm{T}}}}\!\! {{{\left\| {{{\bm{\theta }}_k}\left( {\bf{s}} \right)} \right\|}^2}{\rm{d}}{\bf{s}}}  &\!\approx\! \sum\limits_{\bf{n}}^{\bf N}  {{{\left\| {{{\bf{w}}_{k,{\bf{n}}}}} \right\|}^2}},~\forall k\in\{1,\cdots,K\},
			\end{align}
		\end{subequations}
		where the numbers of reserved expansion items ${\bf N}$ are suggested to be set to
		\begin{align}\label{eqn:NxNyNz}
			{N_x} = \left\lceil {\frac{{{\kappa _0}{L_x}}}{{2\pi }}} \right\rceil ,~{N_y} = \left\lceil {\frac{{{\kappa _0}{L_y}}}{{2\pi }}} \right\rceil ,~{N_z} = \left\lceil {\frac{{{\kappa _0}{L_z}}}{{2\pi }}} \right\rceil.
		\end{align}
	\end{proposition}
	\begin{IEEEproof}
		Employ a truncation operation on (\ref{eqn:Green_Four}) and (\ref{eqn:approx_theta}), and then find the minimum positive integers $N_x$, $N_y$, and $N_z$ subject to $2\pi \frac{{{N_x}}}{{{L_x}}} \ge {\kappa _0}$, $2\pi \frac{{{N_y}}}{{{L_y}}} \ge {\kappa _0}$, and $2\pi \frac{{{N_z}}}{{{L_z}}} \ge {\kappa _0}$.
	\end{IEEEproof}

	Exploiting the above proposition, the total number of expansion items becomes finite, which is acceptable for the practical computing devices. For example, when $L_x=0.5$ m and $f=2.4$ GHz, the number of reserved items on the x-axis can be chosen as $N_x= \left\lceil {\frac{{{\kappa _0}{L_x}}}{2\pi }} \right\rceil =4$. In this way, it becomes feasible to design the continuous pattern functions $\left\{ {{{\bm{\theta }}_k}({\bf{s}})} \right\}_{k = 1}^K$ for \ac{cap-mimo} via optimizing the finite projection lengths $\left\{ {{{\bf{w}}_{k,{\bf{n}}}}} \right\}_{k = 1}^K$ in the wavenumber space.

	\subsection{Performance Analysis of Developed \ac{pdm} Technique}\label{sec:performance_analysis}
	Since the developed \ac{pdm} includes a finite-item approximation as shown in {\it Proposition 1}, the \ac{cap-mimo} system will suffer a performance loss. Nevertheless, it can be expected from {\it Corollary 1} and {\it Corollary 2} that, when the number of the reserved expansion items increases, this performance loss will approach zero asymptotically. In this subsection, to show the \ac{pdm}'s capability of approaching the ideal solution asymptotically, we analyze the performance loss caused by the finite-item approximation. To this end, in order to make the problem analytically tractable and get insightful results, here we consider a simplified \ac{cap-mimo} system with single user ($K=1$), while the general multi-user case will be studied in Section \ref{sec:Method}. 
	
	\subsubsection{Achievable user \ac{snr}}
	To analyze the performance loss caused by the finite-item approximation while employing \ac{pdm}, we first study the achievable user \ac{snr} in the ideal case. Since $K=1$, for notation simplification, we temporarily omit the subscript $k$ of channel function ${{\bf{G}}_k}({\bf{s}})$, pattern function ${{\bm{\theta }}_k}\left( {\bf{s}} \right)$, and polarization combiner $\bm{\psi }_k$, respectively. Then, the original sum-rate maximization problem ${\cal P }_o$ in (\ref{eqn:original_problem}) can be equivalently reformulated as the following \ac{snr} maximization problem:
	\begin{subequations}\label{eqn:snr_max}
		\begin{align}
			{\cal P }_s:~~&\mathop{\max}\limits_{{\bm{\psi }},{{\bm \theta}}\left({\bf s}\right)}~~\gamma  = \frac{{\left| {{{\bm{\psi }}^{\rm{H}}}\int_{{S_{\rm{T}}}}  {\bf{G}}({\bf{s}}){\bm{\theta }}\left( {\bf{s}} \right){\rm{d}}{\bf{s}}} \right|^2}}{{{{\bm{\psi }}^{\rm{H}}}{\bm{\psi }}{\sigma ^2}}}, \\
			&~~{\rm s.t.}~~\int_{{{\cal S}_{\rm{T}}}} {{{\left\| {{\bm{\theta }}\left( {\bf{s}} \right)} \right\|}^2}{\rm{d}}{\bf{s}}}  \le {P_{\rm{T}}},
		\end{align}
	\end{subequations}
	where $\gamma$ is the user \ac{snr}, and ${\bm \psi} \in {\mathbb C}^3$ is used to decode symbols at the user side, as shown in Fig. \ref{img:CAP} (b).
	
	Different from most \ac{cap-mimo} systems whose performance limits are difficult to obtain due to the non-convex functional programming, we prove that the single-user \ac{cap-mimo} system has a closed-form and optimal pattern design, given as below:
	\begin{lemma}[Achievable user \ac{snr}] 
		The maximum achievable \ac{snr} of the single-user \ac{cap-mimo} system, i.e., the optimal solution to problem ${\cal P}_s$ in (\ref{eqn:snr_max}), can be expressed as
		\begin{equation}\label{eqn:optimal_gamma}
			{\gamma ^{{\rm{opt}}}} = \frac{{{P_{\rm{T}}}}}{{{\sigma ^2}}}{\lambda _{\max }}\left\{ {\int_{{{\cal S}_{\rm{T}}}} {{{\bf{G}}}({\bf{s}}){\bf{G}}^{\rm{H}}({\bf{s}}){\rm{d}}{\bf{s}}} } \right\},
		\end{equation}
		where $\lambda_{\max}\left\{\cdot\right\}$ denotes the maximum eigenvalue of its argument. In particular, the optimal user \ac{snr} ${\gamma ^{{\rm{opt}}}}$ can be achieved when 
		\begin{subequations}\label{eqn:single_user_beam}
			\begin{align}
				{\bm{\theta }}\left( {\bf{s}} \right) &\!=\! \frac{\sqrt {{P_{\rm{T}}}}{{{\bf{G}}^{\rm{H}}}({\bf{s}}){{\bm{\xi }}_{\max }}\left\{ {\int_{{S_{\rm{T}}}}\! {{{\bf{G}}}({\bf{s}}'){\bf{G}}^{\rm{H}}({\bf{s}}'){\rm{d}}{\bf{s}}'} } \right\}}}{{\sqrt {\int_{{{\cal S}_{\rm{T}}}} {{{\left\| {{{\bf{G}}^{\rm{H}}}({\bf{s}}){{\bm{\xi }}_{\max }}\left\{ {\int_{{{\cal S}_{\rm{T}}}}\! {{{\bf{G}}}({\bf{s}}'){\bf{G}}^{\rm{H}}({\bf{s}}'){\rm{d}}{\bf{s}}'} } \right\}} \right\|}^2}{\rm{d}}{\bf{s}}} } }},\\
				{\bm{\psi }} &\!=\! \frac{{{\bm{\xi }}_{\max }}\left\{ {\int_{{{\cal S}_{\rm{T}}}} {{{\bf{G}}}({\bf{s}}){\bf{G}}^{\rm{H}}({\bf{s}}){\rm{d}}{\bf{s}}} } \right\}}{\left\|{{\bm{\xi }}_{\max }}\left\{ {\int_{{{\cal S}_{\rm{T}}}} {{{\bf{G}}}({\bf{s}}){\bf{G}}^{\rm{H}}({\bf{s}}){\rm{d}}{\bf{s}}} } \right\}\right\|},
			\end{align}
		\end{subequations}
		where ${\bm \xi}_{\max}\left\{{\cdot}\right\}$ denotes the eigenvector corresponding to the maximum eigenvalue of its argument.
	\end{lemma}
	\begin{IEEEproof}
		Please see Appendix \ref{appendix:singleuser}.
	\end{IEEEproof}
	\begin{remark}
		When the user is located in the far-field region, the squared channel function ${{\bf{G}}^{\rm{H}}}({\bf{s}}){\bf{G}}({\bf{s}})$ takes the similar amplitude for all ${\bf{s}} \in{\cal S}_{\rm T}$, thus we have ${\left\| {{{\bf{G}}^{\rm{H}}}({\bf{s}}){\bf{G}}({\bf{s}})} \right\|} \approx {\left\| {{{\bf{G}}^{\rm{H}}}({\bf{s}}_0){\bf{G}}({\bf{s}}_0)} \right\|}$ wherein ${\bf{s}}_0$ is the central coordinate of ${\cal S}_{\rm T}$. In this case, it can be observed from (\ref{eqn:optimal_gamma}) that, the achievable user \ac{snr} can be approximated by ${\gamma ^{{\rm{opt}}}} \approx \frac{{{P_{\rm{T}}}}}{{{\sigma ^2}}}{A_{\rm{T}}}\left\| {{{\bf{G}}^{\rm{H}}({\bf{s}}_0)}{\bf{G}}({\bf{s}}_0)} \right\| $, where $\left\| {{{\bf{G}}^{\rm{H}}({\bf{s}}_0)}{\bf{G}}({\bf{s}}_0)} \right\|$ can be viewed as the gain of channel between the \ac{cap-mimo} transmitter and the user. One can find that, the achievable user \ac{snr} ${\gamma ^{{\rm{opt}}}}$ is approximately proportional to the transmit \ac{snr} $\frac{{{P_{\rm{T}}}}}{{{\sigma ^2}}}$ and the area of \ac{cap-mimo} aperture $A_{\rm T}$.
	\end{remark}
	
	\subsubsection{Performance loss caused by finite-item approximation}
	By combining {\it Corollary 1} and {\it Lemma 3}, the achievable user \ac{snr} in the ideal case, i.e., $\gamma^{\rm opt}$ in (\ref{eqn:optimal_gamma}), can be equivalently rewritten as 
	\begin{equation}\label{eqn:gamma_e_opt}
		{\gamma ^{{\rm{opt}}}} =  \frac{1}{{{\sigma ^2}}}{\left\| {\sum\limits_{\bf{n}}^\infty  {{{\bf{\Omega }}_{\bf{n}}}{{\bf{w}}_{\bf{n}}}} } \right\|^2},
	\end{equation}
	where ${\bf{\Omega }}_{\bf{n}}$ and ${\bf{w}}_{\bf{n}}$ are the 
	Fourier transform of ${\bf G}\left({\bf s}\right)$ on surface ${\cal S}_{\rm T}$ and the inverse Fourier transform of ${\bm \theta}\left({\bf s}\right)$ at wavenumber ${\bm{\kappa }}_{\bf n}$, respectively, written as
	\begin{align}\label{eqn:Omega_n_w_n}
		{{\bf{\Omega }}_{\bf{n}}} &= \int_{{{\cal S}_{\rm{T}}}} {{\bf{G}}\left( {\bf{s}} \right){\Psi _{\bf{n}}}\left( {\bf{s}} \right){\rm{d}}{\bf{s}}},\\
		{{\bf{w}}_{\bf{n}}} &= \frac{1}{\sqrt{A_{\rm T}}}\int_{{{\cal S}_{\rm{T}}}} {{\bm{\theta }}\left( {\bf{s}} \right)\Psi _{\bf{n}}^*\left( {\bf{s}} \right){\rm{d}}{\bf{s}}}. %= \sqrt {{P_{\rm{T}}}} \frac{{\int_{{{\cal S}_{\rm{T}}}} {{{\bf{G}}^{\rm{H}}}({\bf{s}}){{\bm{\xi }}_{\max }}\left\{ {\int_{{{\cal S}_{\rm{T}}}} {{{\bf{G}}^{\rm{H}}}({\bf{s}}){\bf{G}}({\bf{s}}){\rm{d}}{\bf{s}}} } \right\}\Psi _{\bf{n}}^*\left( {\bf{s}} \right){\rm{d}}{\bf{s}}} }}{{\sqrt {\int_{{{\cal S}_{\rm{T}}}} {{{\left\| {{{\bf{G}}^{\rm{H}}}({\bf{s}}){{\bm{\xi }}_{\max }}\left\{ {\int_{{{\cal S}_{\rm{T}}}} {{{\bf{G}}^{\rm{H}}}({\bf{s}}){\bf{G}}({\bf{s}}){\rm{d}}{\bf{s}}} } \right\}} \right\|}^2}{\rm{d}}{\bf{s}}} } }}
	\end{align}
	After employing the finite-item approximation in {\it Proposition 1} for $\gamma^{\rm opt}$ in (\ref{eqn:gamma_e_opt}), the non-ideal achievable \ac{snr} ${\hat\gamma}$ can be written as:
	\begin{equation}\label{eqn:non-ideal-gamma}
		{\hat\gamma} =  \frac{1}{{{\sigma ^2}}}{\left\| {\sum\limits_{\bf{n}}^{\bf N}  {{{\bf{\Omega }}_{\bf{n}}}{{\bf{w}}_{\bf{n}}}} } \right\|^2}.
	\end{equation}
	To analytically evaluate the truncation error of the finite-item approximation in {\it Proposition 1} on the system performance, here we define a new metric called {\it \ac{snr} loss}: $\Delta:={{{\gamma ^{{\rm{opt}}}} - \hat \gamma } }$, which is related to the \ac{cap-mimo} transmit power and the noise power at the user. Then, our goal is to derive the upper bound of the \ac{snr} loss $\Delta$ for a given $\bf N$, which characterizes the worst-case performance loss caused by the finite-item approximation in {\it Proposition 1}. By exploiting some inequality techniques, we obtain the following lemma:
	\begin{lemma}[Upper bound of \ac{snr} loss] 
		Given the numbers of the reserved Fourier expansion items ${\bf N}=\left(N_x,N_y,N_z\right)$, the \ac{snr} loss $\Delta$ can be upper-bounded by
		\begin{equation}\label{eqn:upper-bound}
			\Delta  \le
			\frac{{{P_{\rm{T}}}}}{{{\sigma ^2}}}\sqrt {1 - \eta } \left( {1 + \sqrt \eta  } \right)\int_{{{\cal S}_{\rm T}}} {\left\| {{\bf{G}}({\bf{s}})} \right\|_{\rm{F}}^2{\rm d}{\bf{s}}},
		\end{equation}
		where $\eta\in[0,1]$ is defined as 
		\begin{equation}
			\eta  = \frac{{\sum\nolimits_{\bf{n}}^{\bf{N}} {\left\| {{{\bf{\Omega }}_{\bf{n}}}} \right\|_{\rm{F}}^2} }}{{\sum\nolimits_{\bf{n}}^\infty  {\left\| {{{\bf{\Omega }}_{\bf{n}}}} \right\|_{\rm{F}}^2} }},
		\end{equation}
		which can be regarded as a threshold describing the degree of finite-item approximation.
	\end{lemma}
	\begin{IEEEproof}
		Please see Appendix \ref{appendix:error}.
	\end{IEEEproof}
	\begin{remark}
		From (\ref{eqn:upper-bound}) we can observe that, the upper bound of the \ac{snr} loss $\Delta$ is approximately proportional to the transmit \ac{snr} $\frac{{{P_{\rm{T}}}}}{{{\sigma ^2}}}$ and the aperture area $A_{\rm T}$ of \ac{cap-mimo}. It implies that, for a given $\bf N$, larger aperture area $A_{\rm T}$ will lead to larger performance loss. This explains why $\bf N$ should be set according to the aperture size, as shown in (\ref{eqn:NxNyNz}). Besides, when the number of the reserved expansion series $\bf N$ increases, i.e., $\eta \to 1$, the upper bound of the \ac{snr} loss $\Delta$ gradually goes to zero. It indicates that, despite the existence of the finite-item approximation, the performance loss can be artificially controlled by setting an acceptable $\bf N$ while employing the developed \ac{pdm}. In this way, the pattern design scheme based on \ac{pdm} is promising to approach the ideal solution asymptotically.
	\end{remark}

	\section{Proposed Pattern Design Scheme Based on \ac{pdm} Technique}\label{sec:Method}
	In this section, to show how to design \ac{cap-mimo} patterns through the developed \ac{pdm} technique, we propose a \ac{bcd} based pattern design scheme to solve the sum-rate maximization problem ${\cal P }_o$ in (\ref{eqn:original_problem}) as a typical example. Specifically, in Subsection \ref{sec:method2}, we present the whole process of the scheme design. Then, in Subsection \ref{sec:method3}, the convergence and computational complexity of the proposed pattern design scheme are discussed.
	
	\subsection{Proposed Pattern Design Scheme for Sum-Rate Maximization}\label{sec:method2}
	In this subsection, we propose a \ac{bcd} based pattern design scheme to solve the sum-rate maximization problem ${\cal P }_o$ in (\ref{eqn:original_problem}). Firstly, we consider to decouple the continuous pattern functions by adopting an equivalent transform for sum-rate maximization problem \cite[\it Theorem 1]{Qingjiang'11}, and we obtain the following lemma.
	\begin{algorithm}[!t] 
		\caption{Proposed pattern design scheme for the sum-rate maximization of multi-user CAP-MIMO systems.} 
		\label{alg:1} 
		\begin{algorithmic}[1] %这个1 表示每一行都显示数字
			\REQUIRE ~~ %算法的输入参数：Input
			Channel functions ${{\bf{G}}_{k}\left({\bf s}\right)}$ with respect to ${\bf s}\in{\cal S}_{\rm T}$ for all users $k\in\{1,\cdots,K\}$.
			\ENSURE ~~ %算法的输出：Output
			Optimized sum-rate $R_{{\rm sum}}$; optimized combiners $\bm{\psi}$ at users; optimized patterns $\bm{\theta}\left({\bf s}\right)$ on the aperture of \ac{cap-mimo} transmitter.
			\STATE Initialize $\bm{\psi}$ and $\bm{\theta}\left({\bf s}\right)$;
			\WHILE {No convergence of $R_{{\rm sum}}$}
			\STATE Update ${\bm \rho}$ by (\ref{eqn:rho_update});
			\STATE Update $\bm{\psi}$ by (\ref{eqn:psi_update});
			\STATE Update ${\bf w}$ by (\ref{eqn:w_update}) and (\ref{eqn:zeta});
			\STATE Update $\bm{\theta}\left({\bf s}\right)$ by (\ref{eqn:w_2_theta});
			\STATE Update $R_{{\rm sum}}$ by (\ref{eqn:objective});
			\ENDWHILE	
			\RETURN Optimized $R^{\rm opt}_{{\rm sum}}$, $\bm{\psi}^{\rm opt}$, and $\bm{\theta}^{\rm opt}\left({\bf s}\right)$. %算法的返回值
		\end{algorithmic}
	\end{algorithm}
	
	\begin{lemma}[Equivalent problem for sum-rate maximization]
		By introducing auxiliary variables ${\bm{\rho }} = {\left[ {{\rho _k}, \cdots ,{\rho _K}} \right]^{\rm{T}}}\in{\mathbb R}^K_{+}$ and the combining vectors ${\bm{\psi }} := \left\{ {{{\bm{\psi }}_k}} \right\}_{k = 1}^K$ for all $K$ users, the original sum-rate maximization problem ${\cal P }_o$ in (\ref{eqn:original_problem}) can be equivalently reformulated as 
		\begin{subequations}\label{eqn:Rsum'}
			\begin{align}
				\!\!\!\!{\cal P }_1:&\mathop{\max}\limits_{{\bm{\rho }}, {\bm \psi}, {{\bm \theta}}\left({\bf s}\right)}~{R_{{\rm{sum}}}'} \!=\! \sum\limits_{k = 1}^K {{{\log }_2}{\rho _k}}  \!-\! \frac{1}{{\ln 2}}\sum\limits_{k = 1}^K {{\rho _k}{E_k}}  \!+\! \frac{K}{{\ln 2}},\\
				&~~~~{\rm s.t.}~~\sum\limits_{k = 1}^K {\int_{{{\cal S}_{\rm{T}}}} {{{\left\| {{{\bm{\theta }}_k}\left( {\bf{s}} \right)} \right\|}^2}{\rm{d}}{\bf{s}}} }  \le {P_{\rm{T}}},
			\end{align}
		\end{subequations}
		where ${\bm \psi}_k\in{\mathbb C}^{3}$ is the polarization combiner at user $k$ as shown in Fig. \ref{img:CAP} (b), and $E_k$ is the mean-square error (MSE) of the decoded symbol ${\hat x}_k={\bm \psi}_k^{\rm H}{\bf y}_k$, defined as
			\begin{align}
				{E_k} =&\, {{\mathbb E}_{{\bf{x}},{\bf{n}}}}\left\{ {\left| {{{\hat x}_k} - {x_k}} \right|}^2 \right\} \notag \\
				= &\,{\left| {1 - \int_{{S_{\rm{T}}}} {{\bm{\psi }}_k^{\rm{H}}{{\bf{G}}_k}({\bf{s}})} {{\bm\theta}_k}\left( {\bf{s}} \right){\rm{d}}{\bf{s}}} \right|^2} + \notag \\ & \sum\limits_{j = 1,j \ne k}^K {{{\left| {\int_{{S_{\rm{T}}}} {{\bm{\psi }}_k^{\rm{H}}{{\bf{G}}_k}({\bf{s}}){{\bm{\theta }}_j}\left( {\bf{s}} \right)} {\rm{d}}{\bf{s}}} \right|}^2}}  + {\sigma ^2}{\left\| {{{\bm{\psi }}_k}} \right\|^2}.
			\end{align}
	\end{lemma}
	
	To solve the equivalent problem ${\cal P }_1$ in (\ref{eqn:Rsum'}), similar to conventional \ac{mimo} beamforming, a scheme of pattern design can be established by optimizing variables $\bm \rho$, combiners $\bm \psi$, and continuous pattern functions ${\bm \theta}({\bf s})$ alternatively until the convergence of sum-rate $R_{\rm sum}$. For clarity, we summarize the whole process of this pattern design scheme in {\bf Algorithm 1}, where the updates of  $\bm \rho$, $\bm \psi$, and ${\bm \theta}({\bf s})$ are introduced in the following three parts, respectively.

	\subsubsection{Fix $\bm \psi$ and ${\bm \theta}({\bf s})$, then optimize ${\bm \rho}$}
	While fixing the user combiners $\bm \psi$ and the \ac{cap-mimo} pattern functions ${\bm \theta}({\bf s})$, the optimal solution to the auxiliary variables ${\bm \rho}$ can be obtained by setting $\frac{{\partial {R'_{\rm sum}}}}{{\partial {\rho _k}}}$ to zero for all $k\in\{1,\cdots,K\}$, written as
	\begin{equation}\label{eqn:rho_update}
		{\rho _k^{\rm opt}} = E_k^{ - 1},~~~k\in\{1,\cdots,K\}.
	\end{equation}
	\subsubsection{Fix ${\bm \rho}$ and ${\bm \theta}({\bf s})$, then optimize $\bm \psi$}
	While fixing the auxiliary variables ${\bm \rho}$ and the pattern functions ${\bm \theta}({\bf s})$ of \ac{cap-mimo}, after removing the unrelated components in problem ${\cal P }_1$ in (\ref{eqn:Rsum'}), the subproblem of optimizing the user combiners ${\bm \psi}$ can be reformulated as
	\begin{align}\label{eqn:subp2}
		{\cal P }_2:\mathop{\max}\limits_{{\bm \psi}}~~ -\sum\limits_{k = 1}^K \rho_k{{\bm{\psi }}_k^{\rm{H}}{\bf{A}}_k{{\bm{\psi }}_k}}  + 2\sum\limits_{k = 1}^K \rho_k{{\mathop{\Re}\nolimits} \left\{ {{\bm{\psi }}_k^{\rm{H}}{{\bm{\beta }}_k}} \right\}},
	\end{align}
	where $\bf{A}_k$ and ${\bm{\beta }}_k$ are defined as
	\begin{subequations}
		\begin{align}
			&{{\bf{A}}_k} = \sum\limits_{j = 1}^K {\int_{{{\cal S}_{\rm{T}}}}\!\!\! {{{\bf{G}}_k}({\bf{s}}){{\bm{\theta }}_j}\left( {\bf{s}} \right)} {\rm{d}}{\bf{s}}{{\left( {\int_{{{\cal S}_{\rm{T}}}}\!\!\! {{{\bf{G}}_k}({{\bf{s}}^\prime }){{\bm{\theta }}_j}\left( {{{\bf{s}}^\prime }} \right)} {\rm{d}}{{\bf{s}}^\prime }} \right)}^{\!\!\rm{H}}}}  \!\!\!+\! {\sigma ^2}{{\bf{I}}_3}, \\
			&{{\bm{\beta }}_k} = \int_{{{\cal S}_{\rm{T}}}} {{{\bf{G}}_k}({\bf{s}}){{\bm{\theta }}_k}\left( {\bf{s}} \right){\rm d}{\bf{s}}}.
		\end{align}
	\end{subequations}
	Note that, subproblem ${\cal P }_2$ in (\ref{eqn:subp2}) is a standard convex \ac{qp}, thus the optimal solution to ${\bm \psi}$ can be easily calculated as
	\begin{align}\label{eqn:psi_update}
		{{\bm{\psi }}_k^{\rm opt}} = {\bf{A}}_k^{ - 1}{{\bm{\beta }}_k},~~~k\in\{1,\cdots,K\}.
	\end{align}
	\subsubsection{Fix ${\bm \rho}$ and $\bm \psi$, then optimize ${\bm \theta}({\bf s})$}
	Given fixed auxiliary variables ${\bm \rho}$ and the user combiners $\bm \psi$, after removing the unrelated components in problem ${\cal P }_1$ in (\ref{eqn:Rsum'}), the subproblem of optimizing the continuous pattern functions ${\bm \theta}({\bf s})$ can be reformulated as
	\begin{subequations}\label{eqn:problem_p3}
		\begin{align}
			\!\!\!\!{\cal P }_3:~~&\mathop{\max}\limits_{{\bm \theta}\left({\bf s}\right)}~~ \sum\limits_{k = 1}^K {\rho_kg_k\left({\bm \theta}\left({\bf s}\right)\right)},\\
			&~~{\rm s.t.}~~\sum\limits_{k = 1}^K {\int_{{{\cal S}_{\rm{T}}}} {{{\left\| {{{\bm{\theta }}_k}\left( {\bf{s}} \right)} \right\|}^2}{\rm{d}}{\bf{s}}} }  \le {P_{\rm{T}}},
		\end{align}
	\end{subequations}
	where the function ${g_k\left({\bm \theta}\left({\bf s}\right)\right)}$ is defined as
	\begin{equation}
		\begin{aligned}
			{g_k}\left( {{\bm{\theta }}\left( {\bf{s}} \right)} \right) =& \sum\limits_{j = 1}^K {{{\left| {\int_{{{\cal S}_{\rm{T}}}} {{\bm{\psi }}_k^{\rm{H}}{{\bf{G}}_k}({\bf{s}}){{\bm{\theta }}_j}\left( {\bf{s}} \right)} {\rm{d}}{\bf{s}}} \right|}^2}}  -\\& 2\,{\mathop{\Re}\nolimits} \left\{ {\int_{{{\cal S}_{\rm{T}}}} {{\bm{\psi }}_k^{\rm{H}}{{\bf{G}}_k}({\bf{s}})} {{\bm{\theta }}_k}\left( {\bf{s}} \right){\rm{d}}{\bf{s}}} \right\}.
		\end{aligned}
	\end{equation}
	To address the challenging functional programming shown in (\ref{eqn:problem_p3}), we consider to employ the continuous-discrete transforms in {\it Corollary 1} and {\it Corollary 2}, as well as the finite-item approximation in {\it Proposition 1}, to address the continuous channel functions and pattern functions. In this way, problem ${\cal P }_3$ in (\ref{eqn:problem_p3}) can be reformulated as	
	\begin{subequations}\label{eqn:problem_p4}
		\begin{align}
			\!\!\!\!{\cal P }_4:~~&\mathop{\max}\limits_{\bf w}~~ \sum\limits_{k = 1}^K {\rho_k{{\hat g}_k}\left( {\bf{w}} \right)},\\
			&~~{\rm s.t.}~~\sum\limits_{k = 1}^K {\sum\limits_{\bf{n}}^{\bf N} {{{\left\| {{{\bf{w}}_{k,{\bf{n}}}}} \right\|}^2}} }  \le {P_{\rm{T}}},
		\end{align}
	\end{subequations}
	where we have defined ${\bf{w}}$ as the set of all projection lengths ${\bf{w}}_{k,{\bf n}}$ and
	\begin{align}
		{{\hat g}_k}\left( {\bf{w}} \right) \!=\! \sum\limits_{j = 1}^K {{{\left| {\sum\limits_{\bf{n}}^{\bf N} {{\bf{h}}_{k,{\bf{n}}}^{\rm{H}}{{\bf{w}}_{j,{\bf{n}}}}} } \right|}^2}}  - 2{\mathop{\Re}\nolimits} \left\{ {\sum\limits_{\bf{n}}^{\bf N} {{\bf{h}}_{k,{\bf{n}}}^{\rm{H}}{{\bf{w}}_{k,{\bf{n}}}}} } \right\},
	\end{align}
	in which ${{\bf{h}}_{k,{\bf{n}}}} := {\bf{\Omega }}_{k,{\bf{n}}}^{\rm{H}}{{\bm{\psi }}_k}$.
	\par
	To simplify notations, we define ${\bf h}_k$  and ${\bf w}_k$ as the vectorized sets of ${\bf h}_{k,{\bf n}}$ and ${\bf w}_{k,{\bf n}}$ for all ${\bf n}=(n_x,n_y,n_z)\in\left\{\{-N_x,\cdots,N_x\},\{-N_y,\cdots,N_y\},\{-N_z,\cdots,N_z\}\right\}$, respectively. In this way, the optimization problem ${\cal P }_4$ in (\ref{eqn:problem_p4}) can be equivalently reorganized as
	\begin{subequations}\label{eqn:problem_p5}
		\begin{align}
			\!\!\!\!{\cal P }_5:~~&\mathop{\max}\limits_{{\bf w}}~~ \sum\limits_{k = 1}^K \rho_k {\left( {\sum\limits_{j = 1}^K {{{\left| {{\bf{h}}_k^{\rm{H}}{{\bf{w}}_j}} \right|}^2}}  \!- 2{\mathop{\Re}\nolimits} \left\{ {{\bf{h}}_k^{\rm{H}}{{\bf{w}}_k}} \right\}} \right)} ,\\
			&~~{\rm s.t.}~~\sum\limits_{k = 1}^K {{{\left\| {{{\bf{w}}_k}} \right\|}^2}}  \le {P_{\rm{T}}}, \label{eqn:power_cons2}
		\end{align}
	\end{subequations}
	which is a standard \ac{qcqp}. By adopting Lagrange multiplier method \cite{admm}, the optimal solution to problem ${\cal P }_5$ in (\ref{eqn:problem_p5}) can be obtained by
	\begin{equation}\label{eqn:w_update}
		\begin{aligned}
			{{\bf{w}}_k^{\rm opt}} = {\rho _k}{\left( {{\rho _k}\sum\limits_{j = 1}^K {{{\bf{h}}_j}{\bf{h}}_j^{\rm{H}} + \zeta {{\bf{I}}_{3{N_F}}}} } \right)^{\!\!- 1}}{{\bf{h}}_k},~ \forall k \in\{1,\cdots,K\},
		\end{aligned}
	\end{equation}
	wherein $N_F:=(2N_x+1)(2N_y+1)(2N_z+1)$ is the total number of the reserved Fourier expansion items. Note that, $\zeta$ is the Lagrange multiplier, which should be chosen such that the complementarity slackness condition of the power constraint (\ref{eqn:power_cons2}) is satisfied, i.e.,
	\begin{equation}\label{eqn:zeta}
		{\zeta ^{\rm opt}} = \min \left\{ {{\zeta} \ge 0:\sum\limits_{k = 1}^K {{{\left\| {{{\bf{w}}_k}} \right\|}^2}}  \le {P_{\rm{T}}}} \right\}.
	\end{equation}
	One-dimensional binary search can be an efficient way to solve (\ref{eqn:zeta}) and obtain the optimal Lagrange multiplier $\zeta^{\rm opt}$ \cite{Qingjiang'11}.
	
	Finally, after calculating the optimal projection lengths $\left\{ {{\bf{w}}_k^{{\rm{opt}}}} \right\}_{k = 1}^K$, according to {\it Lemma 2}, the final solution to the patterns on \ac{cap-mimo} aperture can be obtained by
	\begin{equation}\label{eqn:w_2_theta}
		{\bm \theta}_k^{\rm opt}({\bf s})=\sum\limits_{\bf{n}}^{\bf N}  {{{\bf{w}}^{\rm opt}_{k,{\bf{n}}}}{\Psi_{\bf{n}}}\left( {\bf{s}} \right)},~~{\bf s} \in {\cal S}_{\rm T},
	\end{equation}
	which completes the proposed pattern design scheme.
\begin{remark}
In this paper, we focus on studying the theoretical performance of a multi-user CAP-MIMO system. Due to the difficulties in generating any current distribution and the high-complexity iterative procedure, the online implementation of the proposed pattern design is challenging for current technologies. Therefore, a potential application scenario in the future may be low-mobility communications with slow-fading channels, where the patterns do not need to update frequently. Besides, a possible way to utilize the proposed pattern design is to generate the offline codebook, so that the mobile users can be served through a beam training process.
\end{remark}	
	
\subsection{Convergence and Complexity Analysis}\label{sec:method3}
	%	In this subsection, we analyze the convergence and the complexity of the proposed pattern design scheme, respectively.
	
	\subsubsection{Convergence}
	The convergence of the proposed pattern design scheme is asymptotic due to the finite-item approximation as shown in {\it Proposition 1}. Specifically, here we introduce superscript $t$ as the iteration index for {\bf Algorithm 1}, e.g., ${\bm \theta}^t({\bf s})$ refers to the set of pattern functions at the end of the $t$-th iteration. Thus, the convergence of {\bf Algorithm 1} can be summarized as follows:
	\begin{align}
		&{R_{{\rm{sum}}}'}({\bm{\rho }}^{t+1}, {\bm \psi}^{t+1}, {{\bm \theta}}^{t+1}\left({\bf s}\right))
		\stackrel{(a)}{\geq} {R_{{\rm{sum}}}'}({\bm{\rho }}^{t+1}, {\bm \psi}^{t+1}, {{\bm \theta}}^{t}\left({\bf s}\right))
		\stackrel{(b)}{\geq} \notag \\&~~~
		{R_{{\rm{sum}}}'}({\bm{\rho }}^{t+1}, {\bm \psi}^{t}, {{\bm \theta}}^{t}\left({\bf s}\right))
		\stackrel{(c)}{\geq} {R_{{\rm{sum}}}'}({\bm{\rho }}^{t}, {\bm \psi}^{t}, {{\bm \theta}}^{t}\left({\bf s}\right)),
	\end{align}
	where $(b)$ and $(c)$ follow since the updates of auxiliary variable ${\bm \rho}$ in (\ref{eqn:rho_update}) and combiner $\bm{\psi}$ in (\ref{eqn:psi_update}) are monotonous, while $(a)$ follows when the number of reserved expansion items $N_F$ is sufficiently large. It is because a performance gap exists between problem ${\cal P }_3$ in (\ref{eqn:problem_p3}) and problem ${\cal P }_4$ in (\ref{eqn:problem_p4}), which is caused by the finite-item approximation in (\ref{eqn:truncation}). When $N_F$ increases, this performance gap can be close to zero gradually according to {\it Lemma 4}, which ensures the strict convergence of {\bf Algorithm 1}. % In later simulations, we will show that, for a \ac{cap-mimo} with aperture area of $A_{\rm T}=0.25\,{\rm m}^2$, even when $N_F=9$, the proposed pattern design scheme still has a well performed convergence with each iteration being monotonous.
	\par
	\subsubsection{Complexity}
	%\begin{table}[t]
	%	\centering
	%	\normalsize
	%	\caption{Computational complexity of each iteration step.}
	%	\label{table:1}
	%	\begin{tabular}{|c|c|c|c|c|c|c|}
		%		\hline  % 顶部线
		%		\hline
		%		Variable&Computational complexity ${\cal O}\left(\cdot\right)$\\ 
		%		\hline  % 中部线
		%		${\bm \rho}$& $12K^2I_s+K^2+4K$ \\ \hline
		%		${\bm \psi }$&$10K^2I_s+K^2+21K$ \\ \hline
		%		${\bf w}$& $I_\zeta(N_F^3+2KN_F^2+2K)$ \\ \hline
		%		${\bm \theta}({\bf s})$& $3I_sN_F$\\	
		%		\hline  % 底部线			
		%		\hline
		%	\end{tabular}
	%	\vspace*{-1em}
	%\end{table}
	The computational complexity of the proposed pattern design scheme is mainly introduced by the updates of variables ${\bm \rho}$, ${\bm \psi }$, ${\bf w}$, and pattern functions ${\bm \theta}({\bf s})$, as shown in (\ref{eqn:rho_update}), (\ref{eqn:psi_update}), (\ref{eqn:w_update}), and (\ref{eqn:w_2_theta}), respectively. Let $I_s$ denote the sampling number of the continuous integral operation $\int_{{{\cal S}_{\rm T}}}{\rm d}{\bf s}$. Then, the computational complexity of updating auxiliary variable ${\bm \rho}$ is ${\cal O}\left(12K^2I_s+K^2+4K \right)$, which is mainly caused by the calculation of MSE $E_k$. The complexity of updating combining vector ${\bm \psi }$ is ${\cal O}\left(10K^2I_s+K^2+21K\right)$, which is due to the matrix inversion in (\ref{eqn:psi_update}). Different from the updates
	of ${\bm \rho}$ and ${\bm \psi }$ with closed-form expressions, the update of $\bf w$ requires solving QCQP in (\ref{eqn:problem_p5}). Thus, for a given accuracy tolerance $\varepsilon$, the complexity of updating ${\bf w}$ is ${\cal O}\left( {{{\log }_2}\left( {{1 \mathord{\left/
					{\vphantom {1 \varepsilon }} \right.
					\kern-\nulldelimiterspace} \varepsilon }} \right)\sqrt {{N_F} + 1} (1 + 2{N_F})N_F^3} \right)$, which is caused by the matrix inversion in (\ref{eqn:w_update}) and the one-dimensional binary search for optimal $\zeta^{\rm opt}$. Finally, the complexity of updating ${\bm \theta}({\bf s})$ is ${\cal O}\left(I_sN_F\right)$, which is caused by the discrete-continuous transform as shown in (\ref{eqn:w_2_theta}). In general, since the aperture of \ac{cap-mimo} is nearly continuous, it is reasonable to assume $I_s \gg K$ and $N_F \gg K$. Thus, the overall computational complexity of the proposed pattern design scheme can be approximated by ${\cal O}\left( {{I_o}{I_s}{K^2} + {{\log }_2}\left( {{1 \mathord{\left/
					{\vphantom {1 \varepsilon }} \right.
					\kern-\nulldelimiterspace} \varepsilon }} \right){I_o}N_F^{4.5}} \right)$, wherein $I_o$ is the iteration number required by the convergence of sum-rate $R_{\rm sum}$. Thanks to the finite-item approximation in {\it Proposition 1}, the computational complexity of the proposed pattern design scheme is similar to that of the well-known weighted mean-square error minimization (WMMSE) \cite{Qingjiang'11}.

	\section{Simulation Results}\label{sec:NSR}
	%In this section, we provided extensive simulation results under different conditions to validate the performance of the proposed \ac{pdm} technique.
	
	\subsection{Simulation Setup}
	\begin{figure}[!t]
		\centering
		\includegraphics[width=2.6in]{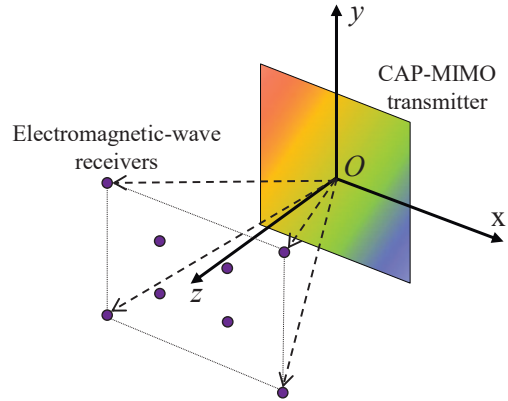}
		%\vspace{-1em}
		\caption{An illustration of the simulation scenario, where one \ac{cap-mimo} transmitter simultaneously serves eight users.}
		\label{img:scenario}
		%\vspace{-1em}
	\end{figure}
	
	\subsubsection{Simulation scenario}
	For the simulation scenario, we consider a 3-D scenario with the topology shown in Fig. \ref{img:scenario}, where one \ac{cap-mimo} transmitter serves $K=8$ users simultaneously. Following the same setup in \cite{Hu'18}, we assume that the \ac{cap-mimo} transmitter is deployed on the $xy$-plane with its center located at $\left(0,0,0\right)$, i.e.,
	\begin{equation}\label{eqn:S_T}
		{{\cal S}_T}: = \left\{ { {\left( {{s_x},{s_y},{s_z}} \right)} {\Big |}\left| {{s_x}} \right| \le \frac{{{L_x}}}{2},\left| {{s_y}} \right| \le \frac{{{L_y}}}{2},{s_z} = 0} \right\},
	\end{equation}
	where the \ac{cap-mimo} aperture has a square shape with the area of $A_{\rm T}=0.25\,{\rm m}^2$, i.e., $L_x=L_y=0.5\,{\rm m}$. Particularly, all users are located in a square region, where four of them are located at $\left(\pm 1\,{\rm m}, \pm 1\, {\rm m}, 30\, {\rm m}\right)$, and the other four are located at $\left(\pm 5\,{\rm m}, \pm 5\,{\rm m}, 30\,{\rm m}\right)$, respectively.
	
	\subsubsection{Simulation parameters}
	Unless otherwise specified, the simulation parameters are set as follows \cite{Fred'08}. The frequency of information-carrying current density ${\bf j}({\bf s})$ and electric field ${\bf e}({\bf r})$ is set to $f=2.4$ $ {\rm GHz}$, and the intrinsic impedance is set to $Z_{0}=376.73$ $\Omega$. The maximum transmit power of \ac{cap-mimo} is set to $P_{\rm T} = 100$ ${\rm mA}^2$ for all schemes to be compared, and the noise power is set to $\sigma^2 = 5.6\times10^{-3}$ ${\rm V}^2/{\rm m}^2$. To observe more space-related insights, channel functions $\left\{ {{{\bf{G}}_k}\left( {\bf{s}} \right)} \right\}_{k = 1}^K$ are generated by the free-space channel model\footnote{ Since the formulations in this paper do not impose requirements on the mathematical structure of channel functions $\{{\bf G}_k({\bf s})\}_{k=1}^K$, the proposed pattern design scheme is also applicable to the other channel models.} (\ref{eqn:free-space}). The sampling number of integral $\int_{{{\cal S}_{\rm T}}}{\rm d}{\bf s}$ is set to $I_s=1024$. To show the impact of finite-item approximation in {\it Proposition 1}, here we consider three different setups for the finite-item approximation of expansion series: (i) $N_x=N_y=7$ and $N_z=0$ (i.e., $N_F=225$); (ii) $N_x=N_y=4$ and $N_z=0$ (i.e., $N_F=81$); (iii) $N_x=N_y=1$ and $N_z=0$ (i.e., $N_F=9$). Note that, for the considered aperture ${\cal S}_{\rm T}$ in (\ref{eqn:S_T}), setup (ii) is exactly the setting of $\bf N$ provided in (\ref{eqn:NxNyNz}), i.e., $\left\lceil {\frac{{{\kappa _0}{L_x}}}{2\pi }} \right\rceil =\left\lceil {\frac{{{\kappa _0}{L_y}}}{2\pi }} \right\rceil=4$. All pattern functions $\left\{ {{{\bm{\theta }}_k}({\bf{s}})} \right\}_{k = 1}^K$ and user combiners $\left\{ {{{\bm{\psi }}_k}} \right\}_{k = 1}^K$ are randomly initialized by the standard complex-Gaussian stochastic processes and variables, respectively. 
	
\subsubsection{Simulation benchmarks}
Inspired by the multi-user pattern designs employed in \cite{Yuan'20} and \cite{Sanguinetti'21}, we consider the following three benchmark schemes for comparison. 

{\bf i) Match-filtering (MF) scheme \cite{Yuan'20}:} Different from our proposed scheme that designs patterns via alternating optimization, the authors in \cite{Yuan'20} directly employed the conjugate of channel functions as the patterns of an uplink multi-user CAP-MIMO system. Although this MF scheme can maximize the desired signal for each user, the inter-user interference cannot be well suppressed. Here we extend this MF scheme to the studied downlink case as a baseline. Without loss of generality, we assume the antennas at users are in the $y$-axis polarization and fix all combiners as ${\bm \psi}_k=\left[0,1,0\right]^{\rm T}$. Then, the pattern functions are set to ${\bm \theta}_k({\bf s}) = \sqrt{p}{\bf G}^{\rm H}_k({\bf s}){\bm \psi}_k$, wherein $p$ is associated with the power allocated to $K$ users. To satisfy the power constraint $\sum\nolimits_{k = 1}^K {\int_{{{\cal S}_{\rm{T}}}} {{{\left\| {{{\bm{\theta }}_k}\left( {\bf{s}} \right)} \right\|}^2}{\rm{d}}{\bf{s}}} }  = {P_{\rm{T}}}$, a scaling operation can be employed to determine the value of $p$. Finally, the optimized patterns ${\bm \theta}_k^{\rm opt}({\bf s})$ can be substituted into (\ref{eqn:SR}) to evaluate the sum-rate.

{\bf ii) Fully-digital MIMO \cite{Sanguinetti'21}:} The authors in \cite{Sanguinetti'21} made a comparison between a CAP-MIMO system and a fully-digital MIMO system by dividing the continuous aperture into several patches spacing of half wavelength. Then, the pattern of each patch is assumed to be rectangular function, of which the amplitude and phase can be optimized like discrete antennas \cite{Sanguinetti'21}. In this paper, we extend this baseline to our studied tri-polarization multi-user case. Specifically, the considered aperture sized of $L_x\times L_y$ allows the fully-digital MIMO to deploy $M = M_x \times M_y = \left\lceil {{2L_x}/{\lambda}} \right\rceil \times \left\lceil {{2L_y}/{\lambda}} \right\rceil$ patch antennas. Assume that each patch antenna has an effective aperture area of $|{{\cal S}_m}|=A_m = \lambda^2/4/\pi$. The $m$-th patch antenna is centered at $(s_{m,x},s_{m,y},0)$ with region ${{\cal S}_m} = \big\{ {( {{s_x},{s_y},{s_z}} )\big|{{\left| {{s_x} - {s_{m,x}}} \right|}^2} + {{| {{s_y} - {s_{m,y}}} |}^2} \le {A_m}/\pi,{s_z} = 0} \big\}$, in which ${s_{m,x}} = {\rm mod}\left(m-1,M_x\right){\lambda}/{2} - {L_x}/{2}$,
${s_{m,y}} = \left\lfloor {{{(m - 1)}}/{{{M_x}}}} \right\rfloor {\lambda}/{2} - {L_y}/{2}$.
Next, the pattern function of the $m$-th patch at region ${\cal S}_m$ is assumed to be ${\bm \theta}_{k,m}({\bf s}) = \frac{1}{\sqrt{A_m}} {{\rm{rect}}\left( {{\bf{s}} \in {{\cal S}_m}} \right){{\bf{v}}_{k,m}}}\in{\mathbb C}^3$, where $\{{\bf{v}}_{k,m}\}_{m=1}^M$ is exactly the digital precoder for user $k$. Therefore, the overall pattern function can be written as 
\begin{equation}\label{eqn:MIMO_pattern}
{{\bm{\theta }}_k}\left( {\bf{s}} \right) = \sum\limits_{m = 1}^M {{\rm{rect}}\left( {{\bf{s}} \in {S_m}} \right){{\bf{v}}_{k,m}}},~k\in\{1,\cdots,K\}.
\end{equation}
By substituting (\ref{eqn:MIMO_pattern}) into the expression of electric fields, we obtain:
\begin{align}
\int_{{{\cal S}_{\rm{T}}}} {{{\bf{G}}_k}({\bf{s}}){{\bm{\theta }}_j}\left( {\bf{s}} \right){\rm{d}}{\bf{s}}}  =& \frac{1}{\sqrt{{A}_m}}\sum\limits_{m = 1}^M {\int_{{{\cal S}_m}} {{{\bf{G}}_k}({\bf{s}}){\rm{d}}{\bf{s}}} {{\bf{v}}_{j,m}}} \notag \\ =& \sum\limits_{m = 1}^M {{{\bf{H}}_{k,m}}{{\bf{v}}_{j,m}}}  = {{\bf{H}}_k}{{\bf{v}}_j}, \label{eqn:MIMO_channel}
\end{align}
where ${{\bf{H}}_{k,m}} = \frac{1}{\sqrt{{A}_m}} \int_{{{\cal S}_m}} {{{\bf{G}}_k}({\bf{s}}){\rm{d}}{\bf{s}}}\in{\mathbb C}^{3\times3}$ is the channel between the $m$-th antenna and user $k$; ${\bf H}_k = \left[ {{{\bf{H}}_{k,1}}, \cdots ,{{\bf{H}}_{k,M}}} \right]$; and ${\bf{v}}_j = \left[{\bf{v}}_{j,m}^{\rm H}, \cdots, {\bf{v}}_{j,M}^{\rm H} \right]^{\rm H}$. By substituting (\ref{eqn:MIMO_pattern}) and (\ref{eqn:MIMO_channel}) into the original problem ${\cal P}_o$ in (\ref{eqn:original_problem}), the sum-rate maximization problem can be reformulated as
	\begin{align}
		\!\!\!\!{\cal P }_f:&\mathop{\max}\limits_{\{{\bf v}_k\}^{K}_{k=1}}~~{R_{\rm sum}} = \sum\limits_{k=1}^K {\log _2}\left(1+ {\bm \alpha}_k^{\rm H}{\bf J}_k^{-1}{\bm \alpha}_k \right) \notag
		\\
		&~~{\rm s.t.}~~\sum\limits_{k = 1}^K {{{\left\| {{{\bf{v}}_k}} \right\|}^2}}  \le {P_{\rm{T}}},
	\end{align}
where ${\bm \alpha}_k \!=\! {{{\bf{H}}_k}{{\bf{v}}_k}}$ and $ {\bf J}_k \!=\! {\sum\nolimits_{j = 1,j \ne k}^K \! {{\bf{H}}_k}{{\bf{v}}_j}{{\left( {{{\bf{H}}_k}{{\bf{v}}_j}} \right)}^{{\rm{H}}}} \!+\! {\sigma ^2}{{\bf{I}}_3}}$. Note that problem ${\cal P }_f$ is exactly a standard sum-rate maximization problem in multi-user MIMO system, which can be solved by the well-known fractional programming in \cite{Shen'18'1}.

{\bf iii) Upper bound:} To evaluate the interference cancellation capability of different schemes, similar to the ideal assumption in multi-user \ac{mimo} systems, we consider the interference-free sum-rate as the upper bound for comparison, which is realized by ideally assuming all inter-user interferences are fully canceled and then employing the proposed pattern design scheme under setup (i) of $\bf N$, i.e., $N_F=225$.
\subsection{Sum-Rate Against the Aperture Area $A_{\rm T}$}
	\begin{figure}[!t]
		\centering
		\includegraphics[width=3.6in]{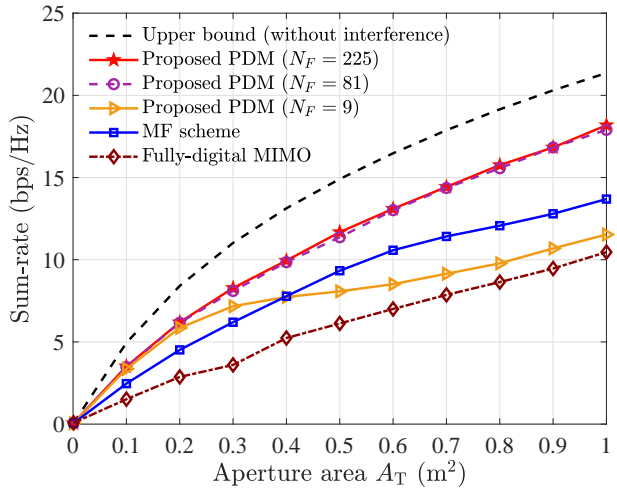}
		%\vspace{-1em}
		\caption{Sum-rate against the aperture area $A_{\rm T}$.}
		\label{img:simulation_size}
		%\vspace{-1em}
	\end{figure}
To show the impact of \ac{cap-mimo} aperture area on the system performance, we first plot the sum-rate against the aperture area $A_{\rm T}$ in Fig. \ref{img:simulation_size}, where the aperture shape of \ac{cap-mimo} always remains a square, i.e., $L_x=L_y$. From this figure, we have the following three observations. 

Firstly, the proposed PDM achieves a higher sum-rate than the MF scheme. For example, when $A_{\rm T}=1\,{\rm m}^2$, the sum-rate achieved by the proposed scheme is 17.90 bps/Hz, which is about 24\% higher than 13.69 bps/Hz achieved by MF scheme. The reason is that, the MF scheme directly employs the conjugate of channel functions to generate $K$ patterns for multiple users, and the negative impact of inter-user interference is ignored. In this way, the beams steering towards different users may spatially overlap, which causes a high interference. In contrast, the proposed PDM jointly designs the patterns for $K$ users, which takes account of the influence of interference in its optimization procedure. During the algorithm implementation, the proposed scheme actually makes a trade-off between the amplification of desired signals and the suppression of inter-user interference. In this way, the EM waves carrying symbols can be accurately steered towards the users with stronger spatial orthogonality, and thus higher sum-rate can be achieved.  
	
Secondly, the proposed \ac{pdm} achieves different performances while given different setups of $\bf N$ for finite-item approximation. Particularly, the \ac{pdm} scheme under setup $N_F=9$ suffers an increasingly large performance loss compared with those under setups $N_F=81$ and $N_F=225$. The reason is that, the reserved number expansion items $N_F=9$ is so small that the finite-item approximation in (\ref{eqn:truncation}) cannot provide enough functional DoFs to well optimize the pattern functions $\left\{ {{{\bm{\theta }}_k}({\bf{s}})} \right\}_{k = 1}^K$. Besides, compared with the \ac{pdm} scheme under $N_F=81$, that under $N_F=225$ achieves almost negligible performance improvement. It implies that, for the considered aperture ${\cal S}_{\rm T}$ in (\ref{eqn:S_T}), when $N_F=81$, the number of functional DoFs has been sufficiently large enough to well design the continuous pattern functions $\left\{ {{{\bm{\theta }}_k}({\bf{s}})} \right\}_{k = 1}^K$. The reason is that, the wavenumber-domain channel gain of \ac{cap-mimo} dominates within the low-wavenumber band of $\left[-\kappa_0,\kappa_0\right]$, as shown in Fig. \ref{img:channel_gain}. Thus, ignoring the high-wavenumber items of Fourier expansion series has very limited impact on pattern designs, which demonstrates the effectiveness of the setup of $\bf N$ provided in (\ref{eqn:NxNyNz}).

Finally, we obtain the similar results in \cite{Sanguinetti'21} that the CAP-MIMO schemes outperform the scheme of fully-digital MIMO. The reason is from three aspects. First, the rectangular function is directly employed as the pattern of each discrete patch. Compared with a continuously controllable patch, each patch of MIMO has limited degree of freedom to manipulate the radiated EM waves. Second, the channel between the $m$-th antenna and user $k$ is ${{\bf{H}}_{k,m}} = \frac{1}{\sqrt{{A}_m}} \int_{{{\cal S}_m}} {{{\bf{G}}_k}({\bf{s}}){\rm{d}}{\bf{s}}}$. Since continuous channel function ${{\bf{G}}_k}({\bf{s}})$ varies at different positions over $\bf{s}\in{{\cal S}_m}$, such an integral may result in a power loss of ${{\bf{G}}_k}({\bf{s}})$ (unless patch is small enough, e.g., $\lambda/10$ diameter \cite{Sanguinetti'21}). Physically speaking, it means that the eigenmode of the patch antenna does not perfectly match the incident waves. That is why smaller reconfigurable antennas are preferred by metasurfaces \cite{badawe2016true}. Third, compared with the practical MIMO with discrete antennas, the reconfigurable aperture of CAP-MIMO fully covers the given region ${\cal S}_{\rm T}$. It determines that CAP-MIMO can obtain higher array gains to enhance the desired signals for users.

\subsection{Sum-Rate Against the Transmit Power $P_{\rm T}$}
	\begin{figure}[!t]
		\centering
		\includegraphics[width=3.6in]{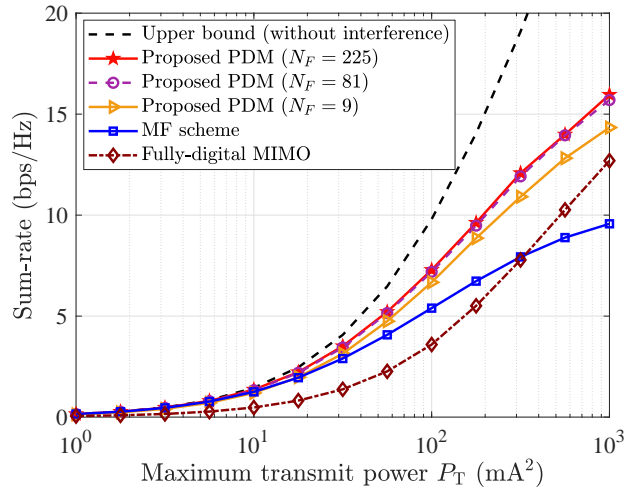}
		%\vspace{-1em}
		\caption{Sum-rate against the maximum transmit power $P_{\rm T}$.}
		\label{img:simulation_PT}
		%\vspace{-1em}
	\end{figure}
	To show the impact of transmit power on the system performance, we plot the sum-rate against the maximum transmit power $P_{\rm T}$ in Fig. \ref{img:simulation_PT}. We have the following third observations. 
	
Firstly, for all considered schemes, the achievable sum-rate increases quickly as the transmit power becomes higher, while the proposed \ac{pdm} always outperforms the benchmark schemes in the considered range of transmit power. For example, when the transmit power $P_{\rm T}$ is $10^{3}\,{\rm{mA}}^2$, the maximum sum-rate achieved by fully-digital MIMO is 12.69 bps/Hz, while that achieved by the proposed \ac{pdm} scheme is about 15.96 bps/Hz, which achieves an improvement of about 26\%.
	
Secondly, the performance gaps, including the gap between the upper bound and the proposed scheme and the gap between the proposed scheme and the MF scheme, become larger as the maximum transmit power $P_{\rm T}$ gets higher. The reason behind this phenomenon is that, the performance gaps among these three schemes highly depend on the inter-user interference, and the sum-rate is simultaneously influenced by the inter-user interference and the noise. When the transmit power is low, the inter-user interference are relatively low, thus the noise dominates, which makes the performance gaps among these schemes small. In contrast, when the transmit power of \ac{cap-mimo} increases, the interference will be more serious, which finally dominates in the undesired factors for sum-rate improvement. 
	
Finally, as $P_{\rm T}$ increases, one can note that fully-digital MIMO outperforms the CAP-MIMO with MF scheme when $P_{\rm T}$ is about $316.2~{\rm{mA}}^2$. The reason is that the MF scheme ignores the inter-user interference. When the transmit power $P_{\rm T}$ rises, the desired signals as well as the uncontrolled interference increase simultaneously, which limits the rate of sum-rate improvement. In contrast, the fraction programming in \cite{Shen'18'1} is adopted to design the precoder of fully-digital MIMO. This scheme actually achieves a balance between amplifying desired signals and suppressing inter-user interference, thus the increase of sum-rate improvement is faster. This indicates that designing interference-suppressed patterns is necessary to make multi-user CAP-MIMO effective.
	
	\begin{figure}[!t]
		\setlength{\abovecaptionskip}{-0.0cm}
		\setlength{\belowcaptionskip}{-0.0cm}
		\centering
		\subfigcapskip -1em
		\subfigure[$k=1$]{
			\includegraphics[width=1.6in]{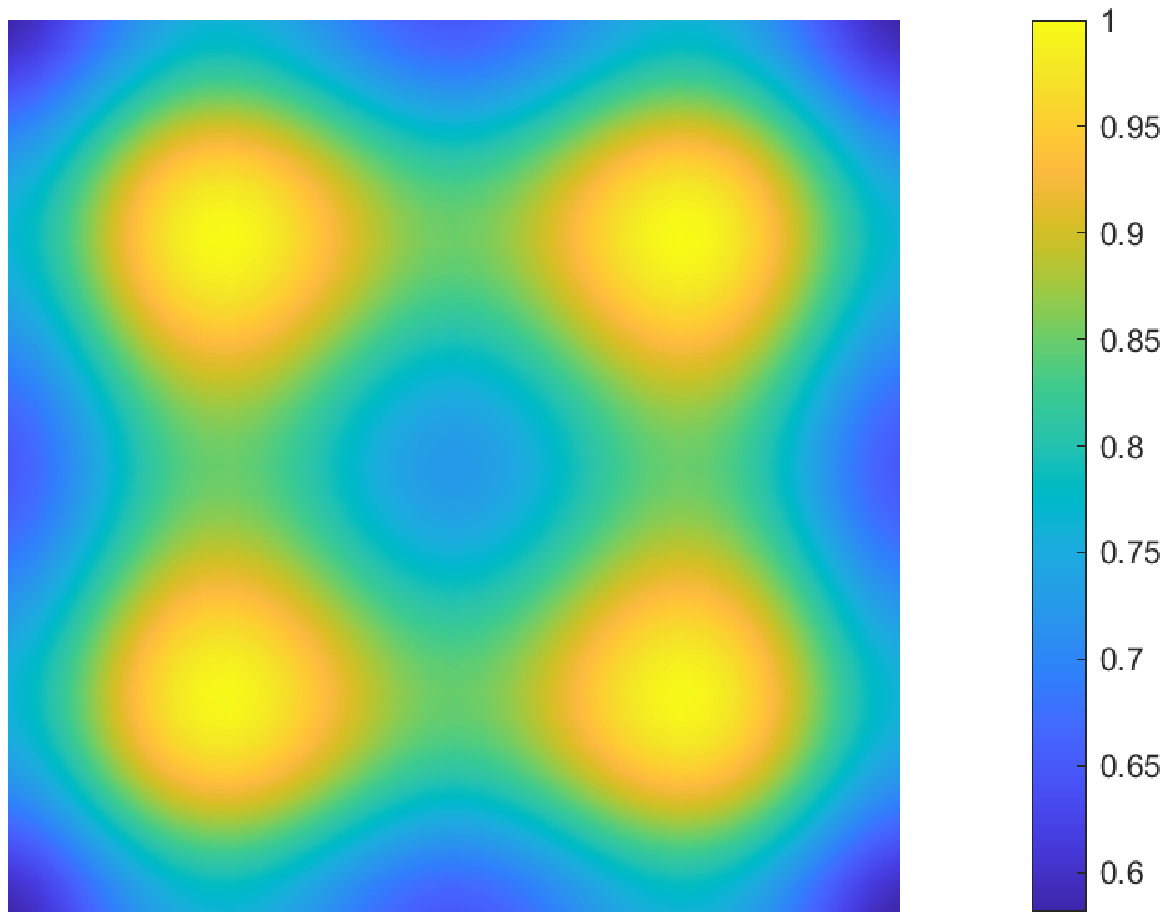}
		}%\hspace{-2em}
		\subfigure[$k=2$]{
			\includegraphics[width=1.6in]{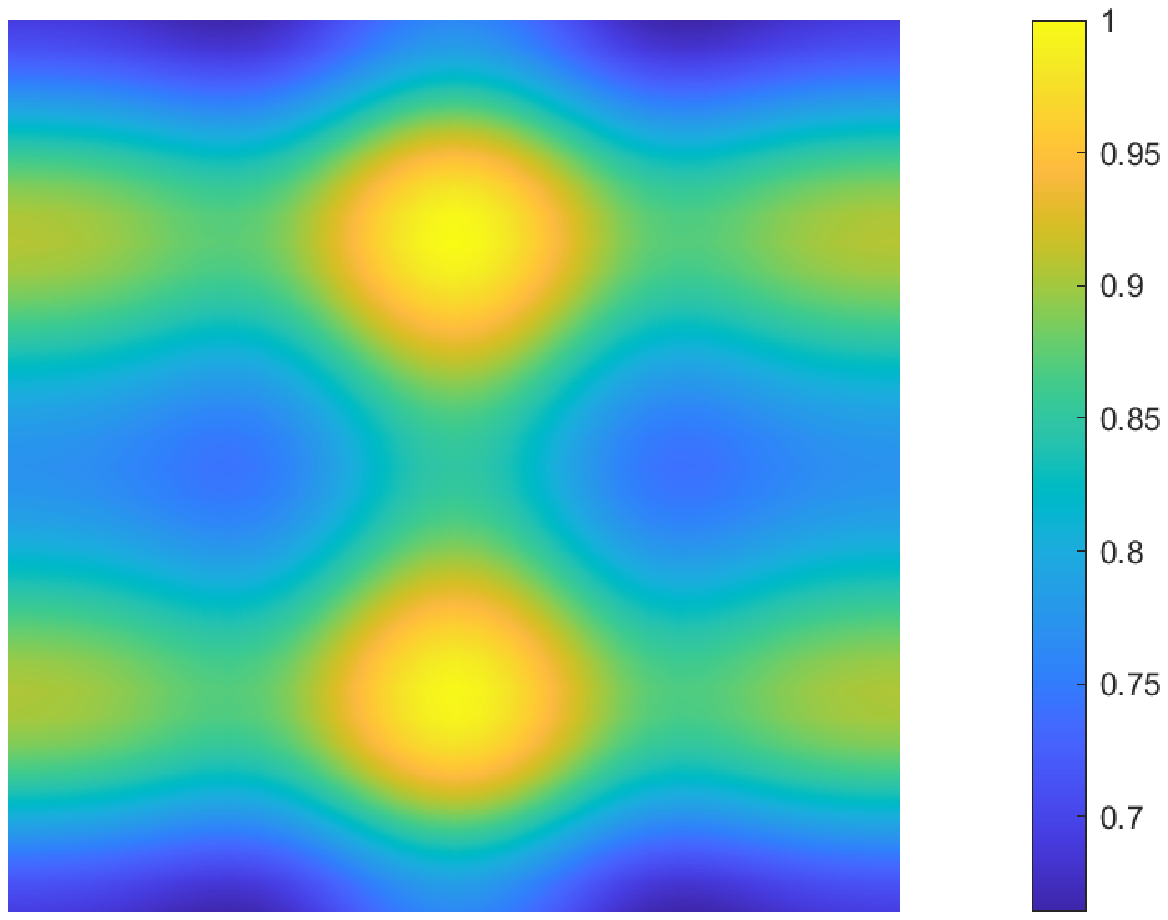}
		}
		%\hspace{-2em}
		\vspace*{-1em}
		\\
		\subfigure[$k=3$]{
			\includegraphics[width=1.6in]{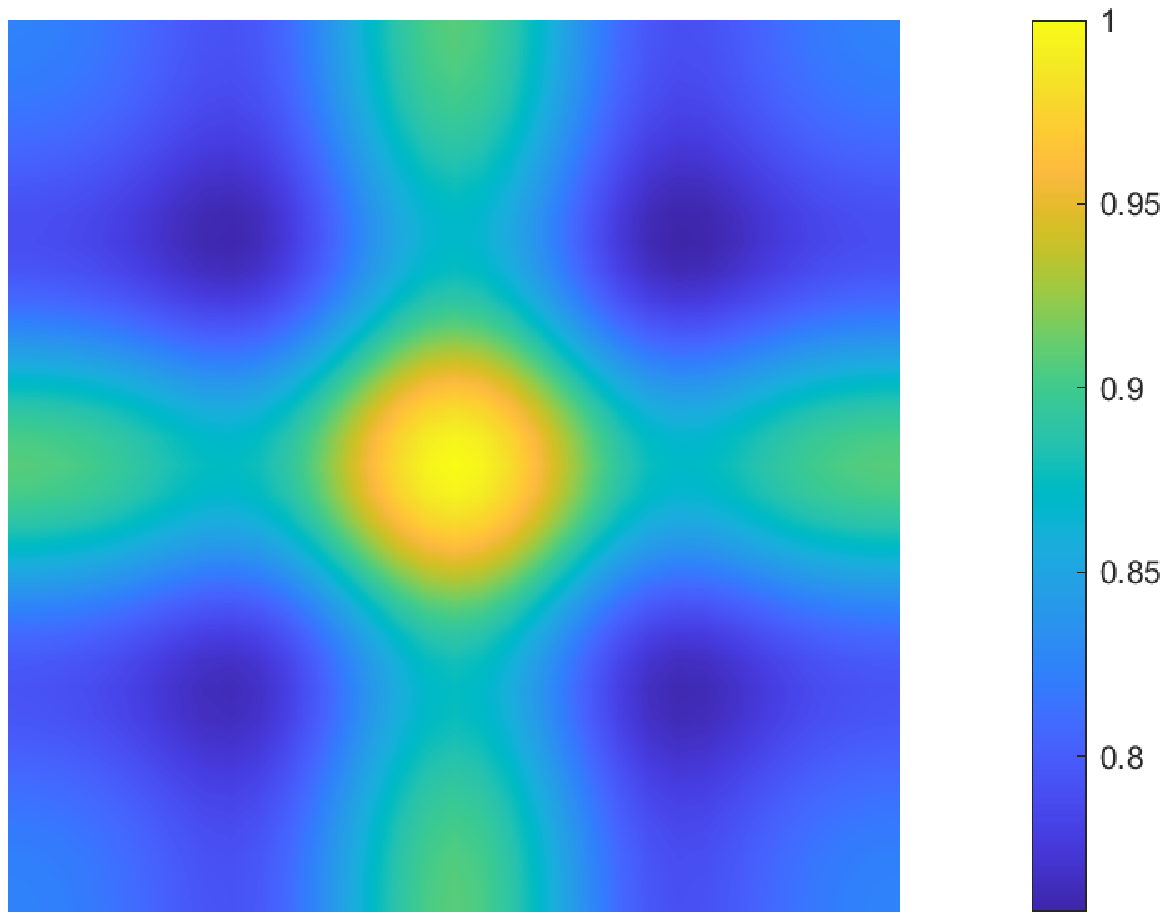}
		}%\hspace{-2em}
		\subfigure[$k=4$]{
			\includegraphics[width=1.6in]{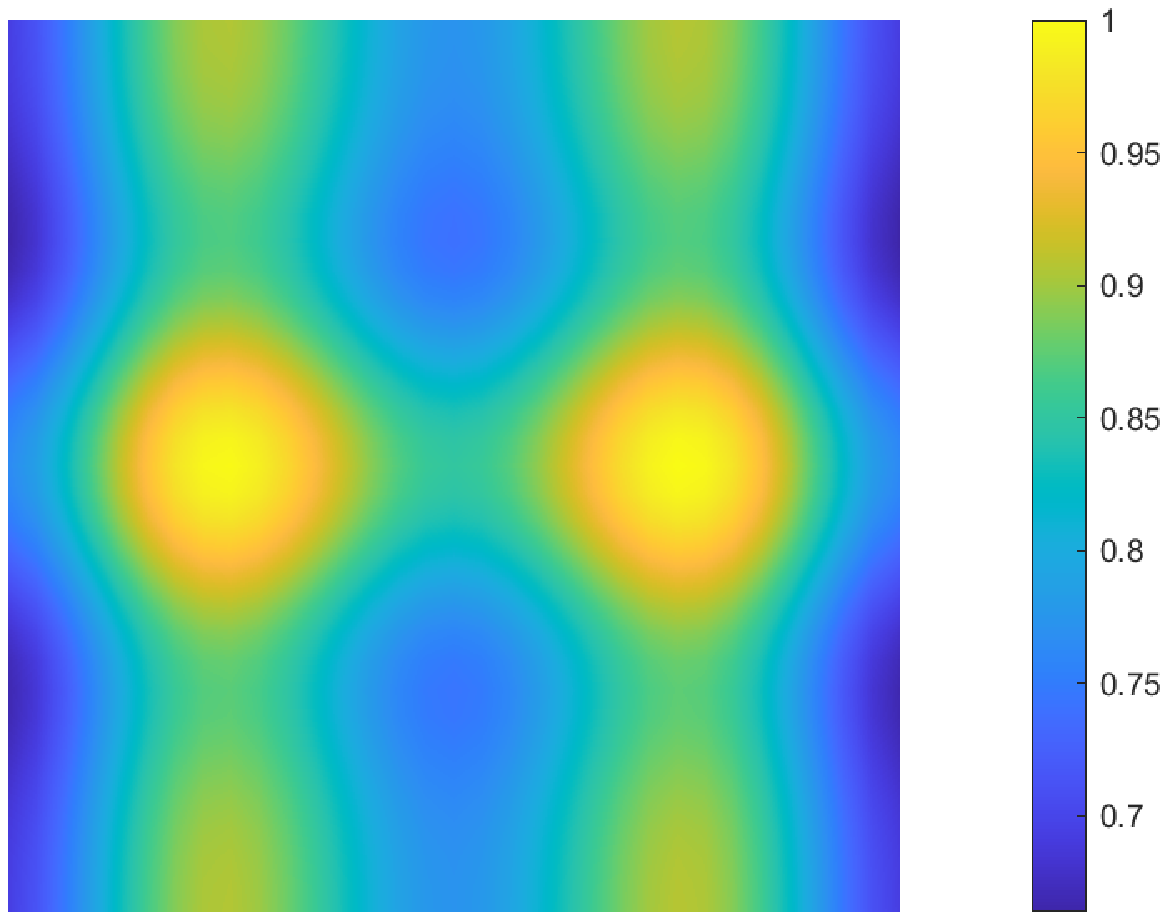}
		}
		%\vspace{1em}
		\caption{
			The normalized amplitude of the x-component of the optimized patterns ${\bm \theta}_k({\bf s})$ with respect to ${\bf s}\in{\cal S}_{\rm T}$ for users $k=$1, 2, 3, and 4.
		}
		%\vspace{-1em}
		\label{img:simulation_amplitude}
	\end{figure}

	\subsection{Patterns ${\bm \theta}({\bf s})$ of \ac{cap-mimo}}

	\begin{figure}[!t]
		\setlength{\abovecaptionskip}{-0.0cm}
		\setlength{\belowcaptionskip}{-0.0cm}
		\centering
		%\hspace{-3em}
		\subfigcapskip -1em
		\subfigure[$k=1$]{
			\includegraphics[width=1.6in]{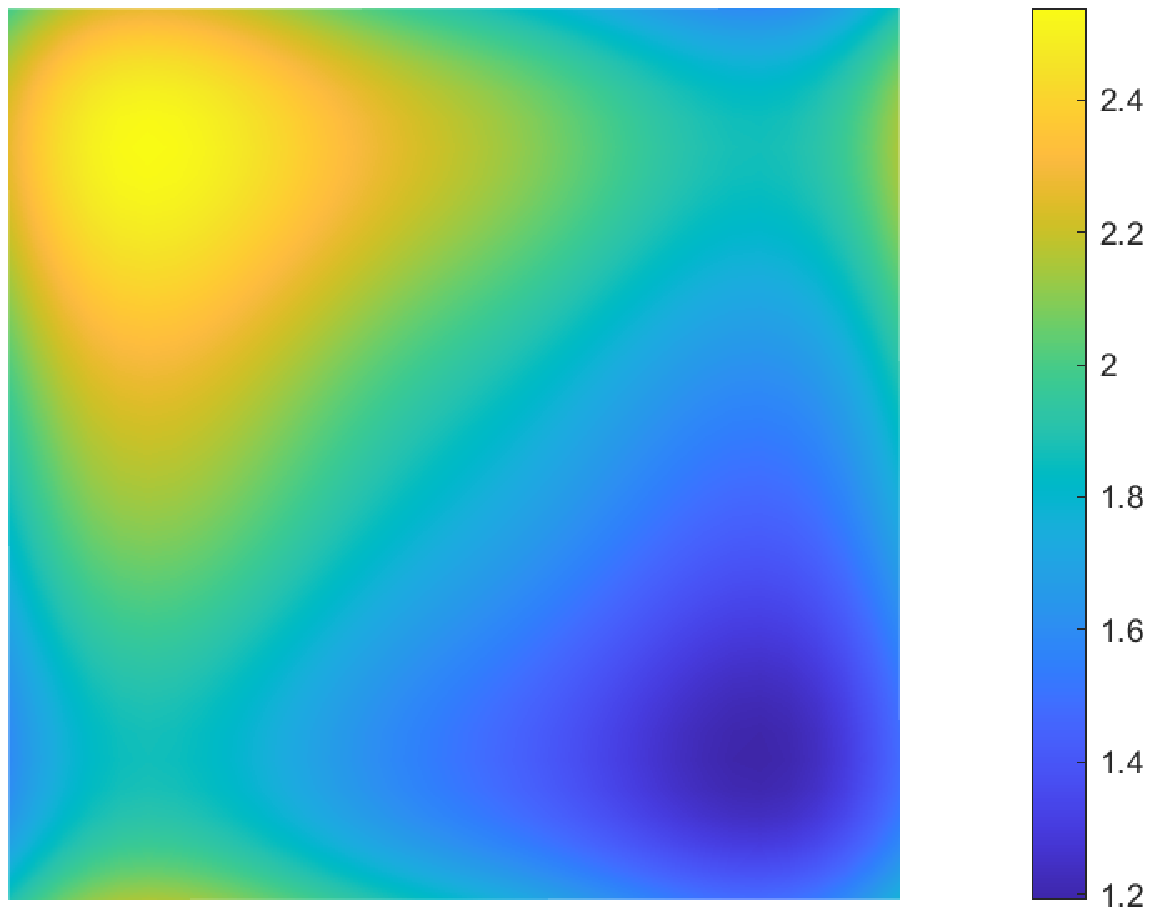}
		}%\hspace{-2em}
		\subfigure[$k=2$]{
			\includegraphics[width=1.6in]{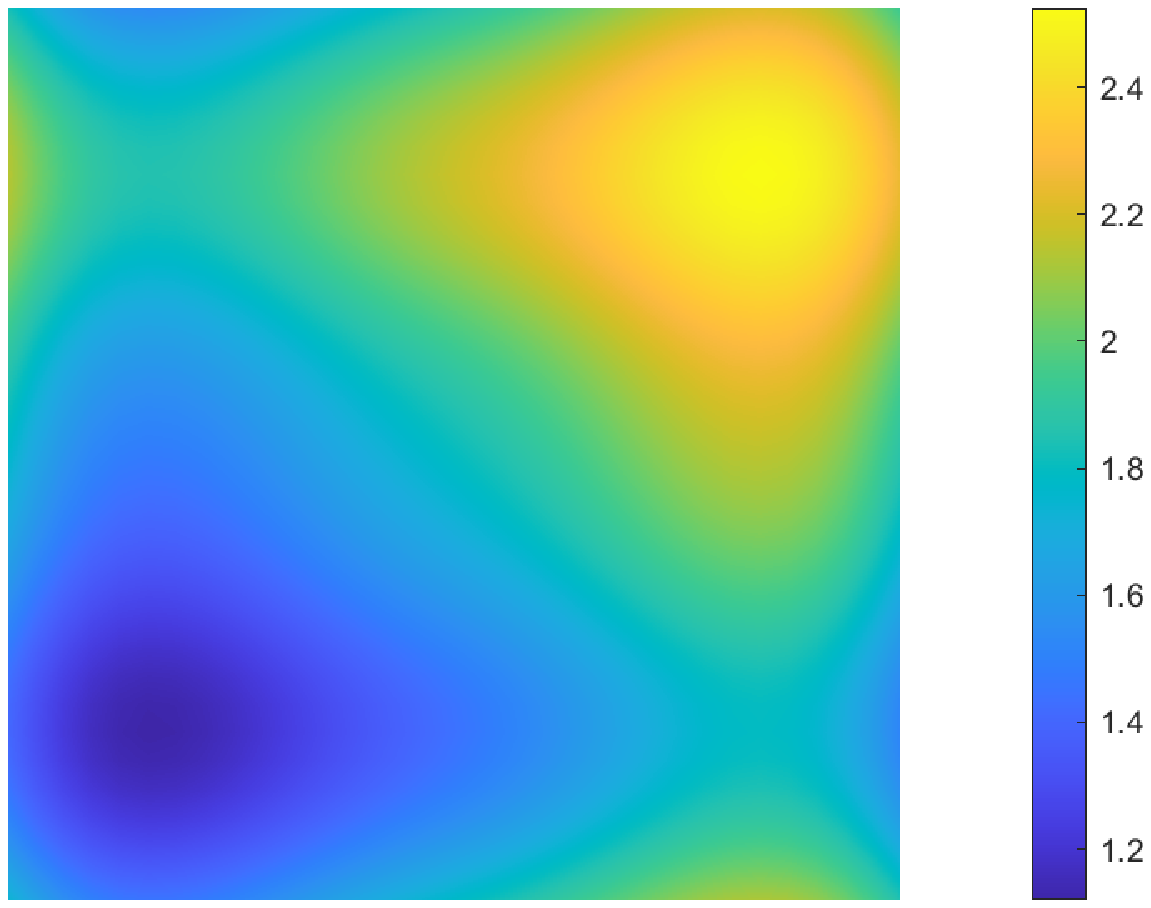}
		}%\hspace{-2em}
		\\
		\vspace*{-1em}
		\subfigure[$k=3$]{
			\includegraphics[width=1.6in]{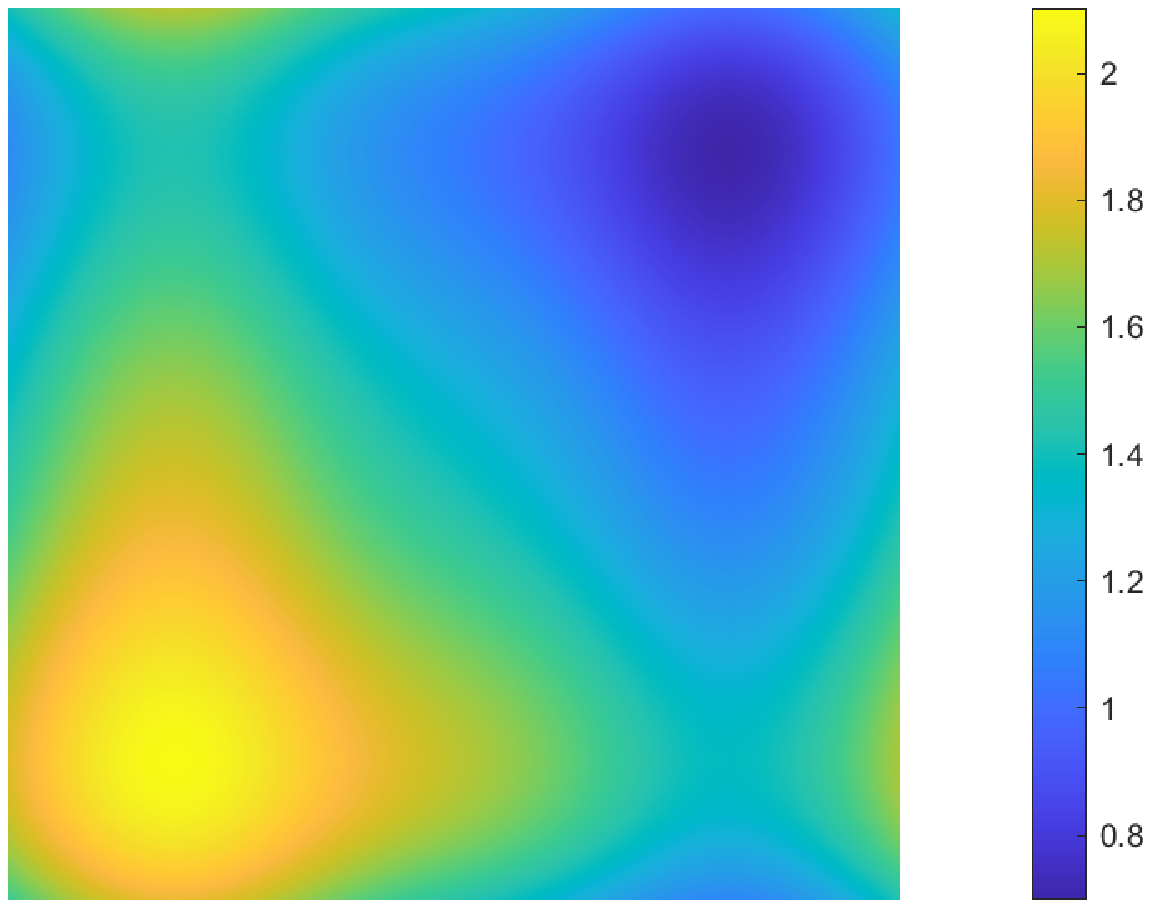}
		}%\hspace{-2em}
		\subfigure[$k=4$]{
			\includegraphics[width=1.6in]{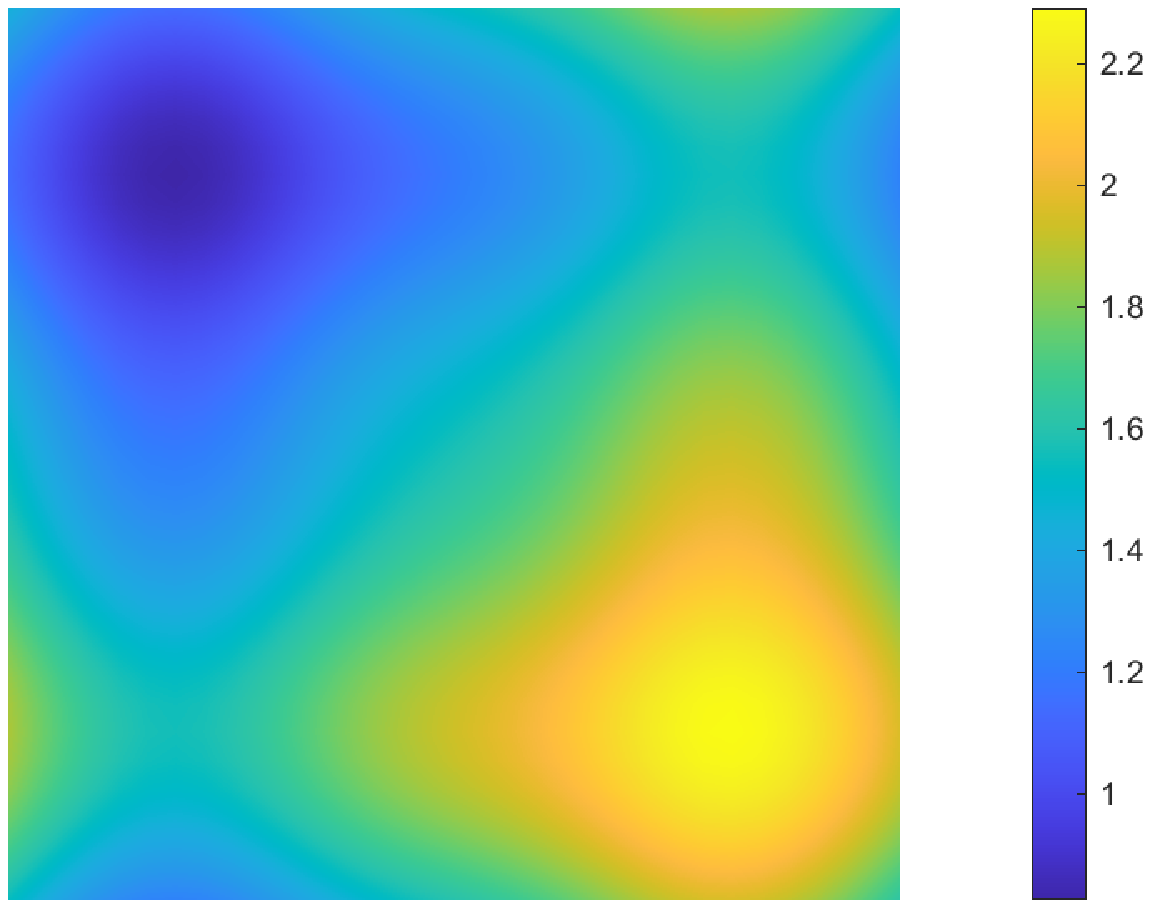}
		}%\hspace{-3em}
		%\vspace{1em}
		\caption{
			The phase of the x-component of the optimized patterns ${\bm \theta}_k({\bf s})$  with respect to ${\bf s}\in{\cal S}_{\rm T}$ for users $k=$1, 2, 3, and 4.
		}
		%\vspace{-1em}
		\label{img:simulation_phase}
	\end{figure}
	To show the pattern functions optimized by the proposed pattern design scheme, by fixing $N_F=81$, we present the normalized amplitude of the x-component of the optimized patterns ${\bm \theta}_k({\bf s})$ for the former four users in Fig. \ref{img:simulation_amplitude}, and their phase in Fig. \ref{img:simulation_phase}, respectively. We observe from Fig. \ref{img:simulation_amplitude} that, after optimizing the patterns via the proposed scheme, the power of the patterns for different users are mainly distributed in non-overlapping regions, which seems like several orthogonal functions. It indicates that, in order to maximize the sum-rate, the pattern functions carrying different symbols are designed to be as orthogonal as possible to reduce the inter-user interference. Besides, from Fig. \ref{img:simulation_phase} one can notice that, the phase values of the patterns carrying different symbols are symmetrically distributed. It indicates that the information-carrying \ac{em} waves are steered toward the angular directions where the four users are located, respectively, as shown in Fig. \ref{img:scenario}. This interesting phenomenon is similar to the result of the conventional \ac{mimo} beamforming, which aims to generate spatially-orthogonal beams toward multiple users. We can conclude that, both of these two figures have provided intuitive explanations for the sum-rate improvement, which has also demonstrated the effectiveness of the proposed pattern design scheme.
	 
	%For example, as shown in Fig. \ref{img:simulation_amplitude}, the power of the patterns carrying different symbols are mainly distributed in non-overlapping regions, in this way, the inter-user interferences at each user can be well eliminated.

\subsection{The Impact of User Positions on Sum-Rate}
\begin{figure}[!t]
	\centering
	\includegraphics[width=3.6in]{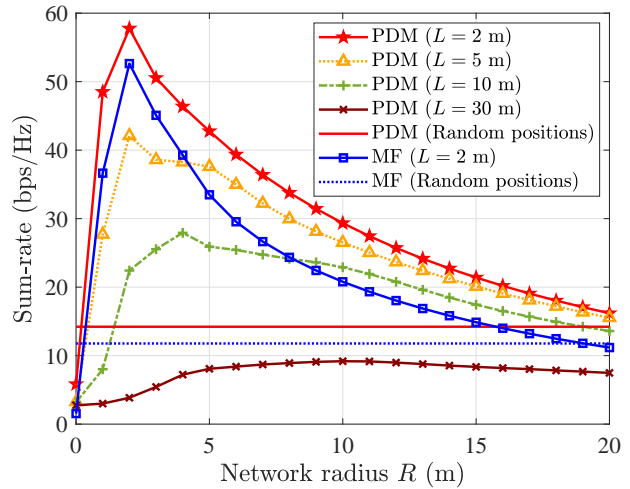}
	%\vspace{-1em}
	\caption{Sum-rate against the network radius $R$.}
	\label{img:simulation_location}
	%\vspace{-1em}
\end{figure}	
In this subsection, we show the impact of users' positions on the sum-rate of a multi-user CAP-MIMO system. Specifically, we consider two different setups. Following the variable-controlling principle, for the first setup, we assume that the distances between all $K$ users and the transmitter's center are equal. Thus, $K$ users are evenly located on a circle centered at $(0,0,L)$ with radius $R$, and the circle is parallel to the $xy$-plane. Particularly, the $k$-th user's position is $\left(R{\rm cos}(\phi_k),R{\rm sin}(\phi_k),L\right)$ wherein $\phi_k = 2k\pi/K$. Typically, we consider four different vertical distances $L=2\,{\rm m}$, $5\,{\rm m}$, $10\,{\rm m}$, $30\,{\rm m}$, and then simulate the sum-rate as a function of network radius $R$. For the second setup, we assume that the positions of $K$ users are randomly distributed in the volume ${\cal V}=\big\{R\in[2\,{\rm m}, 30\,{\rm m}],L\in[2\,{\rm m}, 30\,{\rm m}], \phi_k\in[0,2\pi], \forall k\in\{1,\cdots,K\}\big\}$ to account for the practical case. By fixing $N_F=81$, we plot the sum-rate as a function of network radius $R$ in Fig. \ref{img:simulation_location}, and we obtain the following two observations. 

Firstly, the sum-rates for all schemes decrease as the vertical distance $L$ increases. For example, when $R=10\,{\rm m}$, the proposed PDM scheme can achieve the sum-rate of 29.32 bps/Hz, 26.51 bps/Hz, 22.90 bps/Hz, and 9.18 bps/Hz for $L=2\,{\rm m}$, $5\,{\rm m}$, $10\,{\rm m}$, and $30\,{\rm m}$, respectively. There are two reasons for this phenomenon. One reason is that, as the receivers get far away from the transmitter, more transmitted power is lost into space due to the larger-scale fading of channels, leading to a decrease in the signal strengths at the receivers. The other reason is that, limited by the spatial resolution of the CAP-MIMO aperture, the transmitter is increasingly difficult to accurately steer the desired signals toward the target positions.

%Secondly, as the network radius $R$ becomes larger, the sum-rate of each scheme experiences two stages, i.e., first increases to a peak and then decreases gradually. This interesting phenomenon can be explained as follows. At the first stage, when $R$ is small, the users are so close that the inter-user interference is serious. When $R$ increases, the interference gradually becomes weakened, leading to the increased sum-rate. At the second stage, when $R$ is large enough, the impact of large-scale fading of channels becomes more and more obvious. In this case, the sum-rate starts to decrease.

Secondly, as the network radius $R$ increases, the sum-rate for each scheme experiences two stages: Increase to a peak at first and then gradually decrease at $R=2\,{\rm m}$. This interesting phenomenon can be explained as follows. During the first stage when $R$ is small, the users in the network are close, which results in serious inter-user interference. As $R$ increases, the interference gradually weakens, leading to an increasing sum-rate. Then, during the second stage, when $R$ is sufficiently large, the negative effect of large-scale channel fading becomes more significant, resulting in a decrease of the sum-rate.
%	
%\subsection{Convergence of the Proposed Pattern Design Scheme}
%	\begin{figure}[!t]
%		\centering
%		\includegraphics[width=3.6in]{img/simulation_convergence}
%		\vspace{-1em}
%		\caption{Sum-rate against the iteration number $I_o$.}
%		\label{img:simulation_convergence}
%		\vspace{-1em}
%	\end{figure}
%
%
%
%	To evaluate the convergence performance of the proposed pattern design scheme, in Fig. \ref{img:simulation_convergence}, we plot the sum-rate against the iteration number $I_o$. From this figure, we can find that, the sum-rate achieved by the proposed pattern design scheme is nearly zero when $I_o=1$. 
%	It indicates that, the \ac{em} waves induced by the randomly initialized patterns will mutually cancel at the users, thus they have negligible contribution to sum-rate improvement. However, after several iterations, the sum-rate achieved by the proposed pattern design scheme increases monotonously, even for the scheme under setup $N_F=9$. For example, when $I_o=2$, the sum-rate achieved by the proposed scheme is 5.10 bps/Hz, while that finally converges to about 6.70 bps/Hz when the iteration number $I_o$ is five. We can conclude that, despite the use of finite-item approximation in (\ref{eqn:truncation}), the proposed scheme still enjoys a good convergence performance.
	
\section{Conclusions and Future Works}\label{sec:con}
In this paper, we have developed \ac{pdm} technique to design the \ac{cap-mimo} patterns. Specifically, we have studied and modeled a multi-user CAP-MIMO system, which may provide a framework for some open problems, such as the analyses of channel DoFs and capacity \cite{Zhongzhichao'21}. The developed \ac{pdm} is able to transform the design of the continuous pattern functions to the design of their projection lengths on finite orthogonal bases. Thus, \ac{pdm} may serve as a signal processing framework for some technical problems in CAP-MIMO systems, such as channel estimation \cite{HuangHu'20} and energy efficiency optimization \cite{Huang'18'2}. Utilizing \ac{pdm}, we have proposed a \ac{bcd} based pattern design scheme to solve the formulated sum-rate maximization problem. Simulation results have shown that, the sum-rate achieved by the proposed scheme is higher than that achieved by benchmark schemes. In the future, the CAP-MIMO pattern design with lower complexity is important, and some modern tools like deep reinforcement learning \cite{HuangMo'20} can be leveraged for the efficient pattern design of CAP-MIMO. %{\color{red} By leveraging the property that the channel gain is primarily concentrated within the low-wavenumber band, the compressed-sensing based approach can be explored to reduce the overhead of \ac{cap-mimo} channel estimation.}
	
\appendices
	
	\section{Proof of Lemma 1}\label{appendix:power}
	By integrating the radial component of the Poynting vector over a sphere with infinite-length radius, for a deterministic source in a given surface ${\cal S}_{\rm T}$, the physical radiation power can be upper-bounded by the integral of the Euclidean norm of ${\bf j}({\bf s})$ over ${\bf s}\in{\cal S}_{\rm T}$ \cite{Fred'08}. Here we extend this proof to the case in the sense of expectation, where the current distribution ${\bf j}({\bf s})$ is a stochastic process composed of multiple patterns that carry symbols $\bf x$, as shown in (\ref{eqn:js_theta}). 
	
	Firstly, we calculate the Poynting vector $S({\bf{r}})$ by its definition:
	\begin{align}
		&S({\bf{r}}) \!=\! \frac{1}{{2{Z_0}}}{\bf{e}}^{\rm H}({\bf{r}}){{\bf{e}}}({\bf{r}}) \notag
		\\
		&\!=\!\frac{1}{{2{Z_0}}}\!\sum\limits_{k = 1}^K {{{\left| {{x_k}} \right|}^2}} {\left\| {\int_{{{\cal S}_{\rm{T}}}} {\bf{G}} ({\bf{r}},{\bf{s}}){{\bm{\theta }}_k}\left( {\bf{s}} \right){\rm{d}}{\bf{s}}} \right\|^2} + \notag \\&  \frac{1}{{2{Z_0}}}\!\sum\limits_{j \ne j'}\!{x_j^*{x_{j'}}{{\left( {\int_{{{\cal S}_{\rm{T}}}}\!\!\! {\bf{G}} ({\bf{r}},{\bf{s}}){{\bm{\theta }}_j}\left( {\bf{s}} \right){\rm{d}}{\bf{s}}} \right)}^{\!\!\!\rm{H}}}\! {\int_{{{\cal S}_{\rm{T}}}}\!\!\! {\bf{G}} ({\bf{r}},{\bf{s}}){{\bm{\theta }}_{j'}}\left( {\bf{s}} \right){\rm{d}}{\bf{s}}}},
	\end{align}
	where $\sum\nolimits_{j \ne j'} \triangleq  \sum\nolimits_{j' = 1}^K {\sum\nolimits_{j = 1,j \ne j'}^K {} } $. Utilizing ${\mathbb E}_{\bf x}\{{\bf x}{\bf x}^{\rm H}\}={\bf I}_K$, the physical radiation power of \ac{cap-mimo} ${P_{{\rm{rad}}}}$ in the sense of expectation can be calculated as
	\begin{equation}\label{eqn:Prad}
		\begin{aligned}
			{P_{{\rm{rad}}}} =& \mathop {\lim }\limits_{r \to \infty } {\mathbb E}_{\bf x}\left\{ {\int_{{\Omega }} {S({\bf{r}}){r^2}} {\rm d}{\omega} } \right\}
			\\=& \mathop {\lim }\limits_{r \to \infty } \frac{1}{{2{Z_0}}}\sum\limits_{k = 1}^K {\int_\Omega  {{{\left\| {\int_{{{\cal S}_{\rm{T}}}} \!{\bf{G}} ({\bf{r}},{\bf{s}}){{\bm{\theta }}_k}\left( {\bf{s}} \right){\rm{d}}{\bf{s}}} \right\|}^2}{r^2}} {\rm d}\omega },
		\end{aligned}
	\end{equation}
	wherein $r = \left\| {\bf{r}} \right\|$ and $\omega \in \Omega$ is the solid angle of $4\pi$ steradians. Note that, when $r$ is large enough, the channel function ${\bf{G}} ({\bf{r}},{\bf{s}})$ can be approximated by
	\begin{equation}\label{eqn:approx_G}
		\mathbf{G}(\mathbf{r}, \mathbf{s})=\frac{{{\rm{j}}\kappa_0 {{\rm{Z}}_0}}}{{4\pi }}\frac{{{e^{j\kappa_0 r}}}}{r}\left( {{\bf{I}}_3 - \frac{{{\bf{r}}{{\bf{r}}^{\rm{H}}}}}{{{{r}^2}}}} \right){e^{ - {\rm{j}}{{\bm{\kappa }}^{\rm T}}\left( {\phi ,\varphi }  \right){\bf{s}}}},
	\end{equation}
	where $\phi \in [0, \pi)$ and $\varphi \in [-\pi, \pi)$ are the elevation angle and azimuth angle respectively, which is associated with points $\bf r$ and $\bf s$. Plane-wave wave vector ${\bm{\kappa }}\left( {\phi ,\varphi } \right)$ takes the form of ${\bm{\kappa }}\left( {\phi ,\varphi } \right) = \frac{{2\pi }}{\lambda }{\left[ {\cos \varphi \sin \phi ,\sin \varphi \sin \phi ,\cos \phi } \right]^{\rm{T}}}$. By substituting (\ref{eqn:approx_G}) into (\ref{eqn:Prad}), we obtain
	\begin{align}\label{eqn:Prad_inequality}
		&{P_{{\rm{rad}}}}= \notag\\& \mathop {\lim }\limits_{r \to \infty } \frac{{\kappa _0^2{{\rm{Z}}_0}}}{{32{\pi ^2}}}\sum\limits_{k = 1}^K {\int_\Omega  {{{\left\| {\int_{{{\cal S}_{\rm{T}}}} \!\!{\left( {{{\bf{I}}_3} \!-\! \frac{{{\bf{r}}{{\bf{r}}^{\rm{H}}}}}{{{r^2}}}} \right){e^{ - {\rm{j}}{{\bm{\kappa }}^{\rm{T}}}\left( {\phi ,\varphi } \right){\bf{s}}}}} {{\bm{\theta }}_k}\left( {\bf{s}} \right){\rm{d}}{\bf{s}}} \right\|}^{2}}} \!\!{\rm d}\omega }  \notag \\
		&\!\le\! \mathop {\lim }\limits_{r \to \infty } \frac{{\kappa _0^2{{\rm{Z}}_0}}}{{32{\pi ^2}}}\int_\Omega  {\int_{{{\cal S}_{\rm{T}}}} {{{\left\| {\left( {{{\bf{I}}_3} - \frac{{{\bf{r}}{{\bf{r}}^{\rm{H}}}}}{{{r^2}}}} \right){e^{ - {\rm{j}}{{\bm{\kappa }}^{\rm{T}}}\left( {\phi ,\varphi } \right){\bf{s}}}}} \right\|}^2}} {\rm{d}}{\bf{s}}} {\rm d}\omega  \times \notag \\&~~~~~~~~~ \sum\limits_{k = 1}^K {\int_{{{\cal S}_{\rm{T}}}} {{{\left\| {{{\bm{\theta }}_k}\left( {\bf{s}} \right)} \right\|}^2}} {\rm{d}}{\bf{s}}} 
	\end{align}
	where Cauchy-Schwarz inequality is applied to the right side of the equality to derive its upper bound. It can be observed from (\ref{eqn:Prad_inequality}) that, the component related to $r$ is limited when $r\to \infty$, i.e.,
	\begin{equation}
		\begin{aligned}
			0<\mathop {\lim }\limits_{r \to \infty } \int_\Omega  {\int_{{{\cal S}_{\rm{T}}}} {{{\left\| {\left( {{{\bf{I}}_3} - \frac{{{\bf{r}}{{\bf{r}}^{\rm{H}}}}}{{{r^2}}}} \right){e^{ - {\rm{j}}{{\bm{\kappa }}^{\rm{T}}}\left( {\phi ,\varphi } \right){\bf{s}}}}} \right\|}^2}} {\rm{d}}{\bf{s}}} {\rm d}\omega < \infty.
		\end{aligned}
	\end{equation}
	Therefore, the physical radiation power of \ac{cap-mimo} can be upper-bounded by the component in (\ref{eqn:Prad_inequality}) unrelated to $r$, written as
	\begin{equation}
		\begin{aligned}
			\sum\limits_{k = 1}^K {\int_{{{\cal S}_{\rm{T}}}} {{{\left\| {{{\bm{\theta }}_k}\left( {\bf{s}} \right)} \right\|}^2}{\rm{d}}{\bf{s}}} }  \le {P_{\rm{T}}},
		\end{aligned}
	\end{equation}
	which completes the proof.
	
	\section{Proof of Lemma 3}\label{appendix:singleuser}
	Obviously, the optimal solution to ${\cal P}_s$ in (\ref{eqn:snr_max}) is achieved when $\int_{{{\cal S}_{\rm{T}}}} {{{\left\| {{\bm{\theta }}\left( {\bf{s}} \right)} \right\|}^2}{\rm{d}}{\bf{s}}}  = {P_{\rm{T}}}$. Thus, by applying Cauchy-Schwarz inequality for the objective of problem ${\cal P}_s$ in (\ref{eqn:snr_max}), we obtain 
	\begin{align}
		\gamma &= \frac{{{{\left| {{{\bm{\psi }}^{\rm{H}}}\int_{{{\cal S}_{\rm{T}}}} {\bf{G}} ({\bf{s}}){\bm{\theta }}\left( {\bf{s}} \right){\rm{d}}{\bf{s}}} \right|}^2}}}{{{{\bm{\psi }}^{\rm{H}}}{\bm{\psi }}}{{\sigma ^2}}} \notag \\ &\stackrel{(a)}{\le}  \frac{{\left| {\int_{{{\cal S}_{\rm{T}}}} {{{\left\| {{{\bm{\psi }}^{\rm{H}}}{\bf{G}}({\bf{s}})} \right\|}^2}} {\rm{d}}{\bf{s}}} \right|\left| {\int_{{{\cal S}_{\rm{T}}}} {{{\left\| {{\bm{\theta }}\left( {\bf{s}} \right)} \right\|}^2}} {\rm{d}}{\bf{s}}} \right|}}{{{{\bm{\psi }}^{\rm{H}}}{\bm{\psi }}}{{\sigma ^2}}} \notag \\ & = {P_{\rm{T}}}\frac{{\int_{{{\cal S}_{\rm{T}}}} {{{\left\| {{{\bm{\psi }}^{\rm{H}}}{\bf{G}}({\bf{s}})} \right\|}^2}} {\rm{d}}{\bf{s}}}}{{{{\bm{\psi }}^{\rm{H}}}{\bm{\psi }}}{{\sigma ^2}}},
	\end{align}
	where equality $(a)$ holds when 
	\begin{equation}\label{eqn:theta_appexdix}
		{\bm{\theta }}\left( {\bf{s}} \right) = \sqrt {{P_{\rm{T}}}} \frac{{{{\bf{G}}^{\rm{H}}}({\bf{s}}){\bm{\psi }}}}{{\sqrt {\int_{{{\cal S}_{\rm{T}}}} {{{\left\| {{{\bf{G}}^{\rm{H}}}({\bf{s}}'){\bm{\psi }}} \right\|}^2}{\rm{d}}{\bf{s}}'} } }}
	\end{equation}
	Thus, problem ${\cal P}_s$ in (\ref{eqn:snr_max}) can be equivalently reformulated as
	\begin{align}
		\mathop{\max}\limits_{{\bm{\psi }}}~~\gamma & = {P_{\rm{T}}}\frac{{\int_{{{\cal S}_{\rm{T}}}} {{{\left\| {{{\bm{\psi }}^{\rm{H}}}{\bf{G}}({\bf{s}})} \right\|}^2}} {\rm{d}}{\bf{s}}}}{{{{\bm{\psi }}^{\rm{H}}}{\bm{\psi }}}{{\sigma ^2}}} \notag \\&= {P_{\rm{T}}}\frac{{{{\bm{\psi }}^{\rm{H}}}\left( {\int_{{{\cal S}_{\rm{T}}}} {{{\bf{G}}}({\bf{s}}){\bf{G}}^{\rm{H}}({\bf{s}}){\rm{d}}{\bf{s}}} } \right){\bm{\psi }}}}{{{{\bm{\psi }}^{\rm{H}}}{\bm{\psi }}{\sigma ^2}}}.
	\end{align}
	It is a standard Rayleigh quotient thus $\gamma$ can be maximized when combiner ${\bm{\psi }}$ takes the eigenvector corresponding to the maximum eigenvalue of matrix ${\int_{{{\cal S}_{\rm{T}}}} {{{\bf{G}}}({\bf{s}}){\bf{G}}^{\rm{H}}({\bf{s}}){\rm{d}}{\bf{s}}} }$, written as
	\begin{equation}\label{eqn:xi_opt}
		{\bm{\psi }} = \frac{{{\bm{\xi }}_{\max }}\left\{ {\int_{{{\cal S}_{\rm{T}}}} {{{\bf{G}}}({\bf{s}}){\bf{G}}^{\rm{H}}({\bf{s}}){\rm{d}}{\bf{s}}} } \right\}}{\left\|{{\bm{\xi }}_{\max }}\left\{ {\int_{{{\cal S}_{\rm{T}}}} {{{\bf{G}}}({\bf{s}}){\bf{G}}^{\rm{H}}({\bf{s}}){\rm{d}}{\bf{s}}} } \right\}\right\|}.
	\end{equation}
	Then, the pattern design scheme in (\ref{eqn:single_user_beam}) can be obtained by substituting (\ref{eqn:xi_opt}) into (\ref{eqn:theta_appexdix}) and $\gamma^{\rm opt}$ in (\ref{eqn:optimal_gamma}) can be obtained by substituting (\ref{eqn:single_user_beam}) into (\ref{eqn:snr_max}), which completes the proof.
	
	\section{Proof of Lemma 4}\label{appendix:error}
	For simplifying notations, here we define $\sum\nolimits_{{\bf{n}} = {\bf{N}} + 1}^\infty  {}  = \sum\nolimits_{\bf{n}}^\infty  {}  - \sum\nolimits_{\bf{n}}^{\bf{N}} {}$.
	Then, the full derivation process of {\it Lemma 4} is summarized as follows. 
	
	Firstly, by substituting ${\gamma ^{{\rm{opt}}}}$ in (\ref{eqn:gamma_e_opt}) and $\hat \gamma$ in (\ref{eqn:non-ideal-gamma}) into $\Delta={{{\gamma ^{{\rm{opt}}}} - \hat \gamma } }$, we obtain:
	\begin{align}\label{eqn:appendix_Xi_1}
		&\Delta =\frac{2}{{{\sigma ^2}}}{\Re}\left\{ {{{\left( {\sum\limits_{\bf{n}}^\infty  {{{\bf{\Omega }}_{\bf{n}}}{{\bf{w}}_{\bf{n}}}} } \right)}^{\rm{H}}}\left( {\sum\limits_{{\bf{n}} = {\bf{N}} + 1}^\infty  {{{\bf{\Omega }}_{\bf{n}}}} {{\bf{w}}_{\bf{n}}}} \right)} \right\} - \notag \\&~~~~~~~~~ \frac{1}{{{\sigma ^2}}}{\left\| {\sum\limits_{{\bf{n}} = {\bf{N}} + 1}^\infty  {{{\bf{\Omega }}_{\bf{n}}}} {{\bf{w}}_{\bf{n}}}} \right\|^2}
		\notag \\
		& \stackrel{(a)}{\le}
		\frac{2}{{{\sigma ^2}}}\left\| {\sum\limits_{\bf{n}}^\infty  {{{\bf{\Omega }}_{\bf{n}}}{{\bf{w}}_{\bf{n}}}} } \right\|\left\| {\sum\limits_{{\bf{n}} = {\bf{N}} + 1}^\infty  {{{\bf{\Omega }}_{\bf{n}}}} {{\bf{w}}_{\bf{n}}}} \right\| - \frac{1}{{{\sigma ^2}}}{\left\| {\sum\limits_{{\bf{n}} = {\bf{N}} + 1}^\infty  {{{\bf{\Omega }}_{\bf{n}}}} {{\bf{w}}_{\bf{n}}}} \right\|^2} \notag
		\\
		& \stackrel{(b)}{\le} \frac{1}{{{\sigma ^2}}}\left\| {\sum\limits_{\bf{n}}^\infty  {{{\bf{\Omega }}_{\bf{n}}}{{\bf{w}}_{\bf{n}}}} } \right\|\left\| {\sum\limits_{{\bf{n}} = {\bf{N}} + 1}^\infty  {{{\bf{\Omega }}_{\bf{n}}}} {{\bf{w}}_{\bf{n}}}} \right\| \notag +\\&~~~~~~~~~~~~~~~~~~~~~~ \frac{1}{{{\sigma ^2}}}\left\| {\sum\limits_{\bf{n}}^{\bf{N}} {{{\bf{\Omega }}_{\bf{n}}}{{\bf{w}}_{\bf{n}}}} } \right\|\left\| {\sum\limits_{{\bf{n}} = {\bf{N}} + 1}^\infty  {{{\bf{\Omega }}_{\bf{n}}}} {{\bf{w}}_{\bf{n}}}} \right\|
	\end{align}
	where inequality $(a)$ follows since $\Re\left\{ {{\bf{z}}^{\rm H}{\bf y}} \right\} \le \left\| {\bf{z}} \right\|\left\| {\bf{y}} \right\|$ and $(b)$ follows since $\left\| {\bf{z}} \right\| - \left\| {\bf{y}} \right\| \le \left\| {{\bf{z}} - {\bf{y}}} \right\|$. Then, by substituting ${{\bf{w}}_{\bf{n}}}$ in (\ref{eqn:Omega_n_w_n}) as well as ${\bm{\theta }}({\bf s})$ and ${\bm{\psi }}$ in (\ref{eqn:single_user_beam}) into (\ref{eqn:appendix_Xi_1}), we obtain:
	\begin{align}
		&\Delta	\le
		\frac{1}{{{\sigma ^2}{A_{\rm{T}}}}}\frac{{{P_{\rm{T}}}}}{{\int_{{V_s}} {{{\left\| {{{\bf{G}}^{\rm{H}}}({\bf{s}}){\bm{\psi }}} \right\|}^2}d{\bf{s}}} }}\left\| {\int_{{{\cal S}_{\rm{T}}}} \!{\sum\limits_{\bf{n}}^\infty \! {{{\bf{\Omega }}_{\bf{n}}}} {\Psi _{\bf{n}}}({\bf{s}}){{\bf{G}}^{\rm{H}}}({\bf{s}}){\bm{\psi }}{\rm{d}}{\bf{s}}} } \right\| \notag \\&\times
		\left\| {\int_{{{\cal S}_{\rm{T}}}} {\sum\limits_{{\bf{n}} = {\bf{N}} + 1}^\infty  {{{\bf{\Omega }}_{\bf{n}}}} {\Psi _{\bf{n}}}({\bf{s}}){{\bf{G}}^{\rm{H}}}({\bf{s}}){\bm{\psi }}{\rm{d}}{\bf{s}}} } \right\| + \notag \notag \\ &~~ \frac{1}{{\sigma ^2}{{A_{\rm{T}}}}}\frac{{{P_{\rm{T}}}}}{{\int_{{V_s}} {{{\left\| {{{\bf{G}}^{\rm{H}}}({\bf{s}}){\bm{\psi }}} \right\|}^2}d{\bf{s}}} }}\left\| {\int_{{{\cal S}_{\rm{T}}}} {\sum\limits_{\bf{n}}^{\bf{N}} {{{\bf{\Omega }}_{\bf{n}}}{\Psi _{\bf{n}}}({\bf{s}})} {{\bf{G}}^{\rm{H}}}({\bf{s}}){\bm{\psi }}{\rm{d}}{\bf{s}}} } \right\| \times \notag \\& \left\| {\int_{{{\cal S}_{\rm{T}}}} {\sum\limits_{{\bf{n}} = {\bf{N}} + 1}^\infty  {{{\bf{\Omega }}_{\bf{n}}}} {\Psi _{\bf{n}}}({\bf{s}}){{\bf{G}}^{\rm{H}}}({\bf{s}}){\bm{\psi }}{\rm{d}}{\bf{s}}} } \right\|.
	\end{align}
	Next, by utilizing Cauchy-Schwarz inequality $\left\| {\int_{{{\cal S}_{\rm{T}}}} {{\bf z}^{\rm H}{\bf{y}}{\rm d}{\bf{s}}} } \right\| \le \sqrt {\int_{{{\cal S}_{\rm{T}}}} {{{\left\| {\bf{z}} \right\|}^2}{\rm d}{\bf{s}}} } \sqrt {\int_{{{\cal S}_{\rm{T}}}} {{{\left\| {\bf{y}} \right\|}^2}{\rm d}{\bf{s}}} } $, we further obtain
	\begin{align}
		&\Delta \le \notag \\&
		\frac{{{P_{\rm{T}}}}}{{{\sigma ^2}{A_{\rm{T}}}}}\sqrt {\int_{{{\cal S}_{\rm{T}}}} {{{\left\| {\sum\limits_{\bf{n}}^\infty  {{{\bf{\Omega }}_{\bf{n}}}} {\Psi _{\bf{n}}}({\bf{s}})} \right\|}^2}{\rm{d}}{\bf{s}}} } \sqrt {\int_{{{\cal S}_{\rm{T}}}} {{{\left\| {\sum\limits_{{\bf{n}} = {\bf{N}} + 1}^\infty  \!\!\! {{{\bf{\Omega }}_{\bf{n}}}} {\Psi _{\bf{n}}}({\bf{s}})} \right\|}^2}\!{\rm{d}}{\bf{s}}} }  \notag \\&+ \! \frac{{{P_{\rm{T}}}}}{{{\sigma ^2}{A_{\rm{T}}}}}\sqrt {\int_{{{\cal S}_{\rm{T}}}}\! {{{\left\| {\sum\limits_{\bf{n}}^{\bf{N}} {{{\bf{\Omega }}_{\bf{n}}}{\Psi _{\bf{n}}}({\bf{s}})} } \right\|}^2}\!{\rm{d}}{\bf{s}}} } \sqrt {\int_{{{\cal S}_{\rm{T}}}}\! {{{\left\| {\sum\limits_{{\bf{n}} = {\bf{N}} + 1}^\infty \!\!\! {{{\bf{\Omega }}_{\bf{n}}}} {\Psi _{\bf{n}}}({\bf{s}})} \right\|}^2}\!{\rm{d}}{\bf{s}}} } \notag \\
		&\stackrel{(a)}{\le} \frac{{{P_{\rm{T}}}}}{{{\sigma ^2}}}\sqrt {\sum\limits_{\bf{n}}^\infty  {{{\left\| {{{\bf{\Omega }}_{\bf{n}}}} \right\|}^2}} } \sqrt {\sum\limits_{{\bf{n}} = {\bf{N}} + 1}^\infty  {{{\left\| {{{\bf{\Omega }}_{\bf{n}}}} \right\|}^2}} }  + \notag\\&~~~~~~~~~~~~~~~~ \frac{{{P_{\rm{T}}}}}{{{\sigma ^2}}}\sqrt {\sum\limits_{\bf{n}}^{\bf{N}} {{{\left\| {{{\bf{\Omega }}_{\bf{n}}}} \right\|}^2}} } \sqrt {\sum\limits_{{\bf{n}} = {\bf{N}} + 1}^\infty  {{{\left\| {{{\bf{\Omega }}_{\bf{n}}}} \right\|}^2}} } .
	\end{align}
	where $(a)$ follows since $\left\| {\sum {\bf{z}} } \right\| \le \sum {\left\| {\bf{z}} \right\|}$. Finally, the upper bound of $\Delta$ can be derived as
	\begin{align}
		\Delta \stackrel{(a)}{\le} &
		\frac{{{P_{\rm{T}}}}}{{{\sigma ^2}}}\sqrt {\sum\limits_{\bf{n}}^\infty  {\left\| {{{\bf{\Omega }}_{\bf{n}}}} \right\|_{\rm{F}}^2} } \sqrt {\sum\limits_{{\bf{n}} = {\bf{N}} + 1}^\infty  {\left\| {{{\bf{\Omega }}_{\bf{n}}}} \right\|_{\rm{F}}^2} }  + \notag \\ &~~~~~~~~ \frac{{{P_{\rm{T}}}}}{{{\sigma ^2}}}\sqrt {\sum\limits_{\bf{n}}^{\bf{N}} {\left\| {{{\bf{\Omega }}_{\bf{n}}}} \right\|_{\rm{F}}^2} } \sqrt {\sum\limits_{{\bf{n}} = {\bf{N}} + 1}^\infty  {\left\| {{{\bf{\Omega }}_{\bf{n}}}} \right\|_{\rm{F}}^2} } \notag 
		\\
		\stackrel{(b)}{=}& \frac{{{P_{\rm{T}}}}}{{{\sigma ^2}}}\sqrt {\sum\limits_{\bf{n}}^\infty  {\left\| {{{\bf{\Omega }}_{\bf{n}}}} \right\|_{\rm{F}}^2} } \sqrt {\left( {1 - \eta } \right)\sum\limits_{\bf{n}}^\infty  {\left\| {{{\bf{\Omega }}_{\bf{n}}}} \right\|_{\rm{F}}^2} }  + \notag \\ &~~~~~~~~ \frac{{{P_{\rm{T}}}}}{{{\sigma ^2}}}\sqrt {\eta \sum\limits_{\bf{n}}^\infty  {\left\| {{{\bf{\Omega }}_{\bf{n}}}} \right\|_{\rm{F}}^2} } \sqrt {\left( {1 - \eta } \right)\sum\limits_{\bf{n}}^\infty  {\left\| {{{\bf{\Omega }}_{\bf{n}}}} \right\|_{\rm{F}}^2} }  \notag
		\\
		\stackrel{(c)}{=}& \frac{{{P_{\rm{T}}}}}{{{\sigma ^2}}}\sqrt {1 - \eta } \left( {1 + \sqrt \eta  } \right)\int_{{{\cal S}_{\rm T}}} {\left\| {{\bf{G}}({\bf{s}})} \right\|_{\rm{F}}^2{\rm d}{\bf{s}}}, 
	\end{align}
	where $(a)$ follows since $\left\| {\bf{z}} \right\| \le {\left\| {\bf{z}} \right\|_{\rm{F}}}$; $(b)$ holds since $\eta  = {{\sum\nolimits_{\bf{n}}^{\bf{N}} {\left\| {{{\bf{\Omega }}_{\bf{n}}}} \right\|_{\rm{F}}^2} } \mathord{\left/
			{\vphantom {{\sum\limits_{\bf{n}}^{\bf{N}} {\left\| {{{\bf{\Omega }}_{\bf{n}}}} \right\|_{\rm{F}}^2} } {\sum\limits_{\bf{n}}^\infty  {\left\| {{{\bf{\Omega }}_{\bf{n}}}} \right\|_{\rm{F}}^2} }}} \right.
			\kern-\nulldelimiterspace} {\sum\nolimits_{\bf{n}}^\infty  {\left\| {{{\bf{\Omega }}_{\bf{n}}}} \right\|_{\rm{F}}^2} }}$; and $(c)$ holds according to Parseval's theorem shown in (\ref{eqn:G_Omega}). This completes the proof.
	
\section*{Acknowledgment}
We would like to thank Mr. Zhongzhichao Wan and Mr. Jieao Zhu from Tsinghua University for their professional advice. Their expert knowledge of electromagnetic information theory (EIT) and holographic MIMO (H-MIMO) has helped us a lot to improve the quality of this work.

\footnotesize
%\balance
\bibliographystyle{IEEEtran}
\bibliography{IEEEabrv,reference}
%\balance
\input{biographies}
\end{document}

%% file: biographies.tex
\begin{IEEEbiography}[{\includegraphics[width=1in,height=1.25in,clip,keepaspectratio]{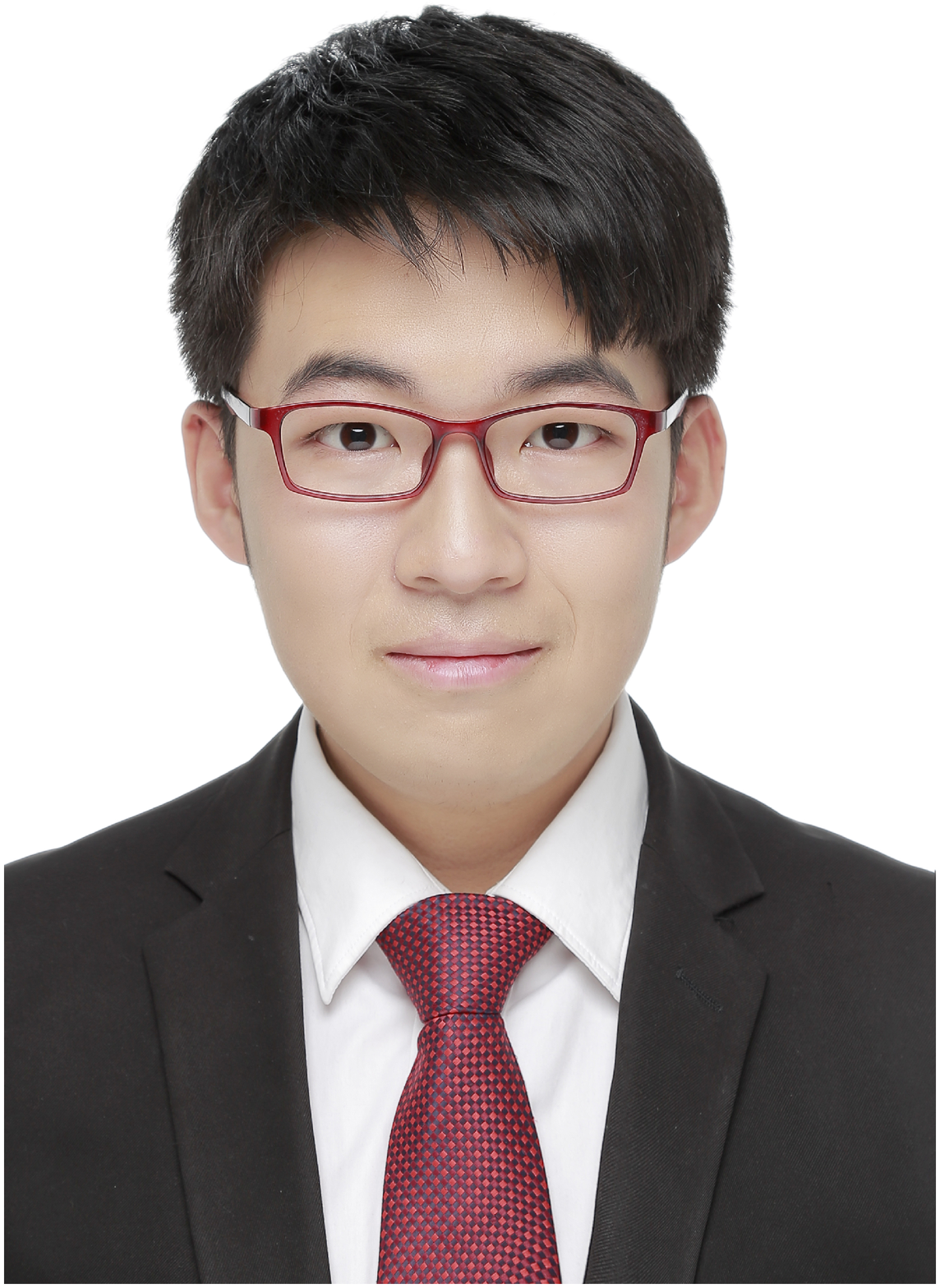}}]{Zijian Zhang}
(Student Member, IEEE) received the B.E. degree in electronic engineering from Tsinghua University, Beijing, China, in 2020. He is currently working toward the Ph.D. degree in electronic engineering from Tsinghua University, Beijing, China.
His research interests include physical-layer algorithms for massive MIMO, holographic MIMO (H-MIMO), and reconfigurable intelligent surfaces (RIS). He is also an amateur in wireless localization and robotics. 

He has authored or coauthored several journal and conference papers for the \textsc{IEEE Journal on Selected Areas in Communications}, the \textsc{IEEE Transactions on Signal Processing}, the \textsc{IEEE Transactions on Wireless Communications}, the \textsc{IEEE Transactions on Communications}, the IEEE ICC, the IEEE GLOBECOM, etc. He has received the National Scholarship in 2019 and the Excellent Thesis Award of Tsinghua University in 2020.

\end{IEEEbiography}

\begin{IEEEbiography}[{\includegraphics[width=1in,height=1.25in,clip,keepaspectratio]{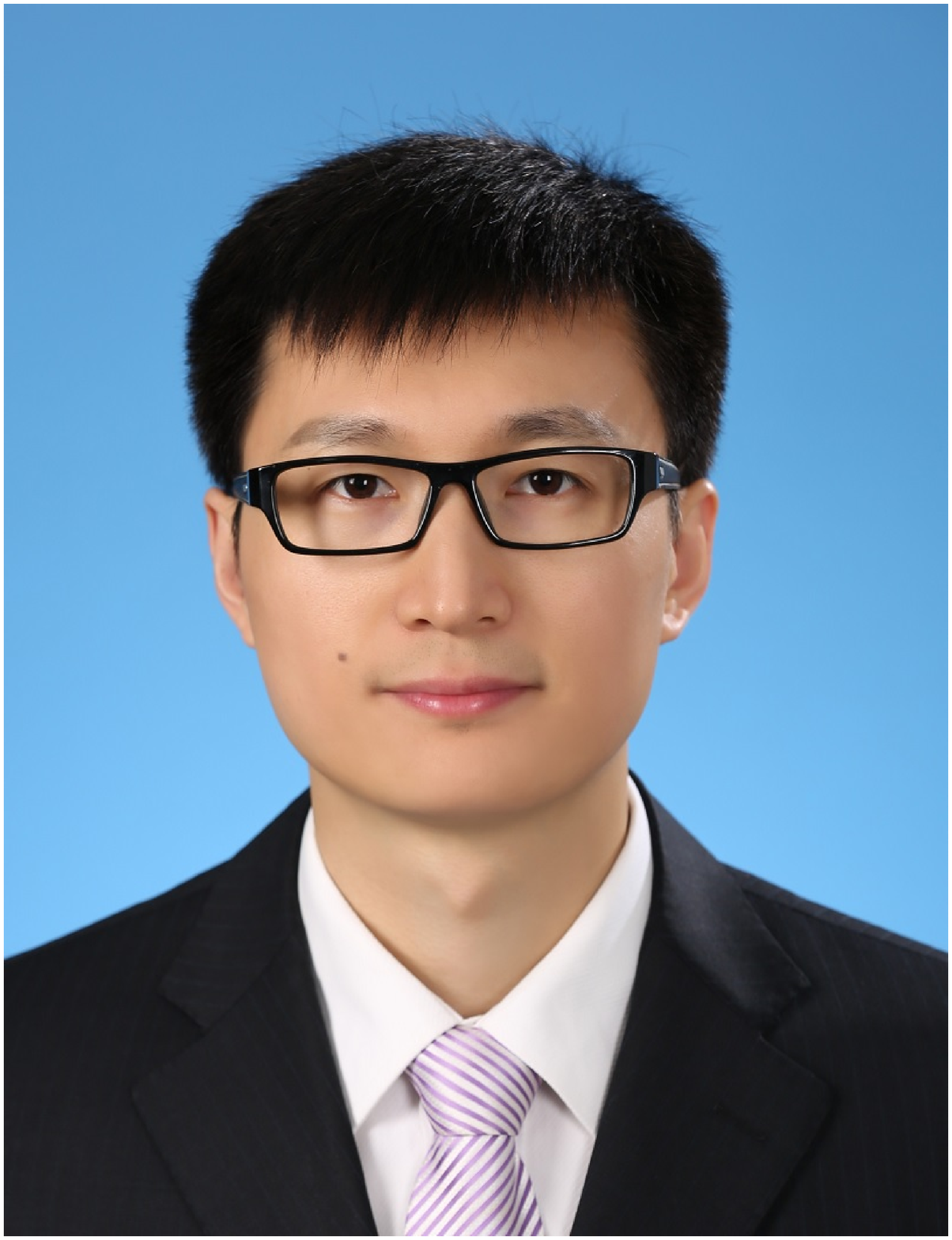}}]{Linglong Dai} (Fellow, IEEE) received the B.S. degree from Zhejiang University, Hangzhou, China, in 2003, the M.S. degree (with the highest honor) from the China Academy of Telecommunications Technology, Beijing, China, in 2006, and the Ph.D. degree (with the highest honor) from Tsinghua University, Beijing, China, in 2011. From 2011 to 2013, he was a Postdoctoral Research Fellow with the Department of Electronic Engineering, Tsinghua University, where he was an Assistant Professor from 2013 to 2016, an Associate Professor since from 2016 to 2022, and has been a Professor since 2022. His current research interests include massive MIMO, reconfigurable intelligent surface (RIS), millimeter-wave and Terahertz communications, machine learning for wireless communications, and electromagnetic information theory. 
	
He has authored or coauthored over 80 IEEE journal articles and over 50 IEEE conference papers. He also holds 19 granted patents. He has coauthored the book {\it MmWave Massive MIMO: A Paradigm for 5G} (Academic Press, 2016). He has received five IEEE Best Paper Awards at the IEEE ICC 2013, the IEEE ICC 2014, the IEEE ICC 2017, the IEEE VTC 2017-Fall, and the IEEE ICC 2018. He has also received the Tsinghua University Outstanding Ph.D. Graduate Award in 2011, the Beijing Excellent Doctoral Dissertation Award in 2012, the China National Excellent Doctoral Dissertation Nomination Award in 2013, the URSI Young Scientist Award in 2014, the IEEE Transactions on Broadcasting Best Paper Award in 2015, the Electronics Letters Best Paper Award in 2016, the National Natural Science Foundation of China for Outstanding Young Scholars in 2017, the IEEE ComSoc Asia Pacific Outstanding Young Researcher Award in 2017, the IEEE ComSoc Asia-Pacific Outstanding Paper Award in 2018, the China Communications Best Paper Award in 2019, IEEE Access Best Multimedia Award in 2020, the IEEE Communications Society Leonard G. Abraham Prize in 2020, and the IEEE ComSoc Stephen O. Rice Prize in 2022, and the IEEE ICC Outstanding Demo Award in 2022. He was listed as a Highly Cited Researcher by Clarivate Analytics from 2020 to 2022. He was elevated as an IEEE Fellow in 2022.
	
He is currently serving as an Area Editor of the \textsc{IEEE Communications Letters}, and an Editor of the \textsc{IEEE Transactions on Wireless Communications}. He has also served as an Editor of the \textsc{IEEE Transactions on Communications} (2017-2021), an Editor of the \textsc{IEEE Transactions on Vehicular Technology} (2016-2020), and an Editor of the \textsc{IEEE Communications Letters} (2016-2020). He has also served as a Guest Editor of the \textsc{IEEE Journal on Selected Areas in Communications}, \textsc{IEEE Journal of Selected Topics in Signal Processing}, \textsc{IEEE Wireless Communications}, etc. 
	
Particularly, he is dedicated to reproducible research and has made a large amount of simulation code publicly available.
	
\end{IEEEbiography}